\documentclass[twocolumn]{aastex63}

\pdfoutput=1 
\usepackage{amsmath,amstext}
\usepackage[T1]{fontenc}
\usepackage{apjfonts} 
\usepackage[figure,figure*]{hypcap}
\usepackage{booktabs}
\usepackage{multirow}
\usepackage{lipsum}
\usepackage{comment}
\usepackage{textcomp}
\usepackage{hyperref}
\usepackage{rotating}
\usepackage{tikz}
\usepackage{float}
\usepackage{graphicx}
\usepackage[caption=false]{subfig}
\usepackage{supertabular}
\usepackage{subfig}
\usepackage[english]{babel}
\usepackage{tablefootnote}

\newcommand{\eg}{e.g., }
\newcommand{\ie}{i.e., }

\newcommand{\rev}[0]{{\bf }}  

\newcommand{\revi}[0]{{\bf }}  

\newcommand{\revii}[0]{{\bf }}  

\received{xxx, xxx}
\revised{xxx, xxx}
\accepted{xxx}
\submitjournal{AJ}

\shorttitle{Nauyaca: a new tool to determine planetary masses and orbital elements}
\shortauthors{Canul et al.}

\begin{document}

\title{\texttt{Nauyaca}: a new tool to determine planetary masses and orbital elements through transit timing analysis}


\correspondingauthor{Eliab Canul}
\email{ecanul@astro.unam.mx}

\author[0000-0002-6185-2886]{Eliab F. Canul}
\affiliation{\rev{Universidad Nacional Aut\'onoma de M\'exico, Instituto de Astronom\'ia, AP 70-264, CDMX  04510, M\'exico}}

\author[0000-0002-5181-4528]{H\'ector Vel\'azquez}
\affiliation{\rev{Universidad Nacional Aut\'onoma de M\'exico, Instituto de Astronom\'ia, AP 70-264, CDMX  04510, M\'exico}}

\author[0000-0002-7486-6726]{Yilen G\'omez Maqueo Chew}
\affiliation{\rev{Universidad Nacional Aut\'onoma de M\'exico, Instituto de Astronom\'ia, AP 70-264, CDMX  04510, M\'exico}}


\begin{abstract}

Transit Timing Variations (TTVs) \revi{is} currently the most successful method to determine dynamical masses and orbital elements for Earth-sized transiting planets. Precise mass determination is fundamental to restrict planetary densities and thus infer planetary compositions. 
In this work, we present \texttt{Nauyaca}, a Python package dedicated to find planetary masses and orbital elements through the fitting of observed mid-transit times from a N-body approach.
The fitting strategy consists in performing a sequence of minimization algorithms (optimizers) that are used to identify high probability regions in the parameter space. These results from optimizers are \revi{used for initialization of} a Markov \revi{chain} Monte Carlo (MCMC) method, using an adaptive Parallel-Tempering algorithm. A set of runs are performed in order to obtain posterior distributions of planetary \revi{masses and orbital elements.} 
In order to test the tool, we created a mock catalog of synthetic planetary systems with different number of planets where all of them transit. We  calculate their mid-transit times to give them as an input to \texttt{Nauyaca}, testing statistically its efficiency in recovering the planetary parameters from the catalog.
\revi{For the recovered planets, we find typical dispersions around the real values of $\sim$1-14 $\mathrm{M_{\oplus}}$  for masses, between 10-110 seconds for periods and  between $\sim$0.01-0.03 for eccentricities.
We also investigate the effects of the signal-to-noise and number of transits in the correct determination of the planetary parameters. Finally, we suggest choices of the parameters that govern the tool, for the usage with real planets,} according to the \revi{complexity of the problem} and computational facilities.

\end{abstract}



\section{Introduction} \label{sec:intro}

Transit timing variations \citep[TTVs; ][]{2005Sci...307.1288H,2005MNRAS.359..567A}  is to date the most successful method to measure precise masses of Earth-sized transiting planets harbored in multiplanet systems that could not be identified by other means, for instance, radial velocities \citep[][]{2016MNRAS.457.4384S, 2017ApJ...839L...8M}.
TTVs contain valuable information that allows us to determine orbital properties which are useful to predict transit ephemeris, orbital stability and dynamical evolution \citep[for a review of TTVs see][and references therein]{2018haex.bookE...7A}.

Deriving planet parameters (masses and orbital elements) from observed transit times is known as the TTVs inversion problem. Since the first planetary system was characterized by TTVs \citep[Kepler-9;][]{2010Sci...330...51H} many authors have used TTVs to determine planetary masses for systems where all known planets transit \citep[\eg][]{Masuda2014ApJ...783...53M,2020MNRAS.491.5238S,2020arXiv201001074A} and also to characterize non-transiting planets
\cite[\eg][]{2012Sci...336.1133N,2013ApJ...777....3N,2015ApJ...812L..18B,2017AJ....153..198S,  2017AJ....154...64M,2018A&A...620A..88C}.

Two main approaches have been largely developed to invert TTVs based on analytical and numerical approximations.
Analytical approaches take advantage of the low computational cost at the expense of the limitation to specific scenarios such as planets near mean motion resonances, low eccentric orbits or specific two-planet systems \citep[][]{2012ApJ...761..122L, 2018ApJ...860...16L,2008ApJ...688..636N,2009ApJ...701.1116N,2010ApJ...709L..44N,2016ApJ...818..177A}.
Numerical models including N-body integrations seem to be an unavoidable route to study more diverse scenarios than those considered by the analytical techniques but also to complement and double-check results from these analytical methods.

The inversion problem requires methods to fully explore the parameter space of the planets independently of the type of models mentioned above. Many works have used combinations of techniques and models to deal with the TTVs inversion problem.
For example, minimization routines and analytical models \citep[][]{2010ApJ...709L..44N}; \rev{genetic algorithm and numerical N-body models \citep{2018A&A...620A..88C}; a coupled simulated annealing algorithm plus N-body \citep{2010ApJ...718..543M}}; Markov \revi{chain} Monte Carlo (MCMC) methods and analytic models \citep[][]{2019MNRAS.484.3772T}; MCMC and N-body \citep[][]{2015ApJ...812L..18B,Jontof_Hutter_2016,2020arXiv201001074A}; minimization plus MCMC using N-body simulations \citep{Masuda2014ApJ...783...53M};
\rev{multimodal nested sampling algorithm combined with either N-body  \citep[\eg][]{2013ApJ...777....3N,2017AJ....153..198S,2020MNRAS.491.5238S}, or with both, N-body and analytic models \citep[\eg][]{2017AJ....154...64M,2021ApJ...908..114Y};
}
a combination of MCMC plus analytical and numerical models \citep[][]{2012ApJ...761..122L,2016ApJ...828...44H,2017AJ....154....5H}, and MCMC and minimization plus analytical tools \citep[][]{2019OEJV..197...71G}. 
Despite of the numerous works that employ computational tools, there is a scarcity of available tools in a ready-to-use package that allows to deal with the TTVs inversion problem in an intuitive, easy and confident way.

In this work, we introduce \texttt{Nauyaca}\footnote{\textit{Bothrops asper}, a kind of pit viper from Central America \citep{Heath2005}}, an easy-to-use Python tool dedicated to deal with the TTVs inversion problem from the N-body approach. The numerical tool, \revi{even being} computational expensive compared to analytical approximations, is more adaptable to address general situations as, for example, the number of planets, prograde or retrograde orbits, planets out of resonances, with eccentric and non-coplanar orbits.
The only required data are transit ephemeris per planet and the stellar mass and radius. Additionally, any prior knowledge about any planetary parameter can be supplied in order to better constrain the parameter space.

This paper is structured as follows: in Section \ref{sec:methods}, we describe the main features of the tool and the modules functionality. Section \ref{sec:catalog} describes the creation of a mock catalog of synthetic planetary systems and mid-transit times to test the tool. In section \ref{sec:settings} we discuss the election of the parameter space and the parameters for the minimization and MCMC algorithms.
In section \ref{sec:Rec_results}, we apply \texttt{Nauyaca} to the whole simulated catalog and discuss the consistency between the input planetary parameters and those found by the tool. Section \ref{caveats} is dedicated to briefly highlight caveats of our tool, and finally in section \ref{sec:Summary} we summarize our findings and we make suggestions about the procedure and parameters to deal with real data.

\section{Methods} \label{sec:methods}

We implemented a Python package, named \texttt{Nauyaca}, focused in the determination of planet masses and orbital elements through mid-transit times fitting for planets around single parent stars. Our tool is equipped with minimization routines and a Markov \revi{chain} Monte Carlo method exclusively adapted to fit transit times series from an N-body approach. 
\revi{\texttt{Nauyaca} manages the exploration of the parameter space with the main goal of finding solutions of planetary parameters that produce mid-transit times consistent with observations, for each planet.}
The tool can work in a parallelized scheme, so multi-core machines are preferred for best performing.

We incorporated \texttt{TTVFast}\footnote{Original C version: \url{https://github.com/kdeck/TTVFast}. A Python wrapper around C version: \url{https://github.com/mindriot101/ttvfast-python}} \citep{2014ApJ...787..132D} an optimized fast code (5-20 times faster than standard methods) \revi{to make transit timing models.}  
\revi{In short, \texttt{TTVFast} receives a set of initial conditions for the planets (mass and orbits) and the stellar mass, and perform an N-body simulation. At the same time, an incorpored Keplerian interpolator calculates  mid-transit times by interpolating the orbits when a planet is detected crossing the star.}
Even though N-body simulations could be time consuming and computationally expensive, we decided to not implement analytical or semi-analytical model approximations because many of these models are just valid  for planets in low order eccentricities, near first-order orbital resonances or valid for a fixed number of planets. We opted for a general purpose approach that could work with more diverse planetary configurations and thus, we decided to use \texttt{TTVFast}.

We define $\Theta_j \equiv \{\mathrm{mass},\ P,\ ecc,\ inc,\ \omega,\ M,\ \Omega \}_j$ as a set of planet parameters \revi{for the j-th planet, where the parameters are, respectively:} the mass, period, eccentricity, inclination, argument of periastron, mean anomaly and longitude of ascending node.
\revi{The orbital elements in $\Theta_j$ correspond to the instantaneous elements at a specific reference time $t_0$ which is specified by the user or automatically selected by \texttt{Nauyaca}. 
Orbital elements are defined in a fixed astrocentric coordinate system with the X-Y plane spanning the plane of the sky and the observer located orthogonally at +Z.}

\revi{Given a constant stellar mass $M_*$ and radius $R_*$, the algorithms incorporated in \texttt{Nauyaca}  make proposals to conform an initial condition ($\Theta$) for all the planets in order to run \texttt{TTVFast}, which results in a model of transit times per planet. Then, to perform the TTVs fitting}, we assume that transit errors are independent, following a \revi{normal} distribution, and thus we set the log-likelihood function in the form,

\begin{equation}
    \log \mathcal{L}_j(\Theta_j|\mathrm{T}_j)\ =\ -\frac{1}{2}\chi^{2}_j -\ \sum_{i}^{N_j}\  \frac{1}{2}\ \log (2\pi \sigma^2_{j,i})
    \label{Loglikelihood}
\end{equation}

\noindent where,

\begin{equation}
    \chi^2_j =  \sum_{i}^{{\mathrm{N_{tran}}}}\ \left[ \frac{t_{j}(i)\ -\ t_{j}^{sim}(i) }{\sigma_j (i)}  \right]^2,
    \label{eq_chi2}
\end{equation}

\noindent with $j$ denoting numbered planets and $i$ denoting their respective transit epochs over the total number of  transits N$_{\mathrm{tran}}$.  Here $\mathit{t}$ corresponds to the observed transit time for a specified epoch,  \revi{ $t^{\textit{sim}}$ is the simulated transit time calculated by the model and $\sigma$ corresponds to the uncertainty in the central time}. We take errors $\sigma_{j}(i)$ as the mean of the upper and lower errors of the $i$-th transit time. \revi{The total log-likelihood is calculated by adding the individual log-likelihoods $\log \mathcal{L}_j$ of the planets.}

\rev{The fitting procedure is performed over the available transit times, but it is possible to include planets in the system without transit times that interact with the transiting planets. The explored space also includes those of the non-transiting (or undetected) planets, and thus we can get information about their planetary parameters. However, considering planets without transit data is left for a future work.}

\subsection{Modules}

\texttt{Nauyaca} incorporates several modules with techniques adapted specifically for the TTVs inversion problem. We describe the functionalities of these modules below.

\subsubsection{Setup} \label{subsec:setup}

\begin{figure}
    \centering
    \includegraphics[scale=0.52]{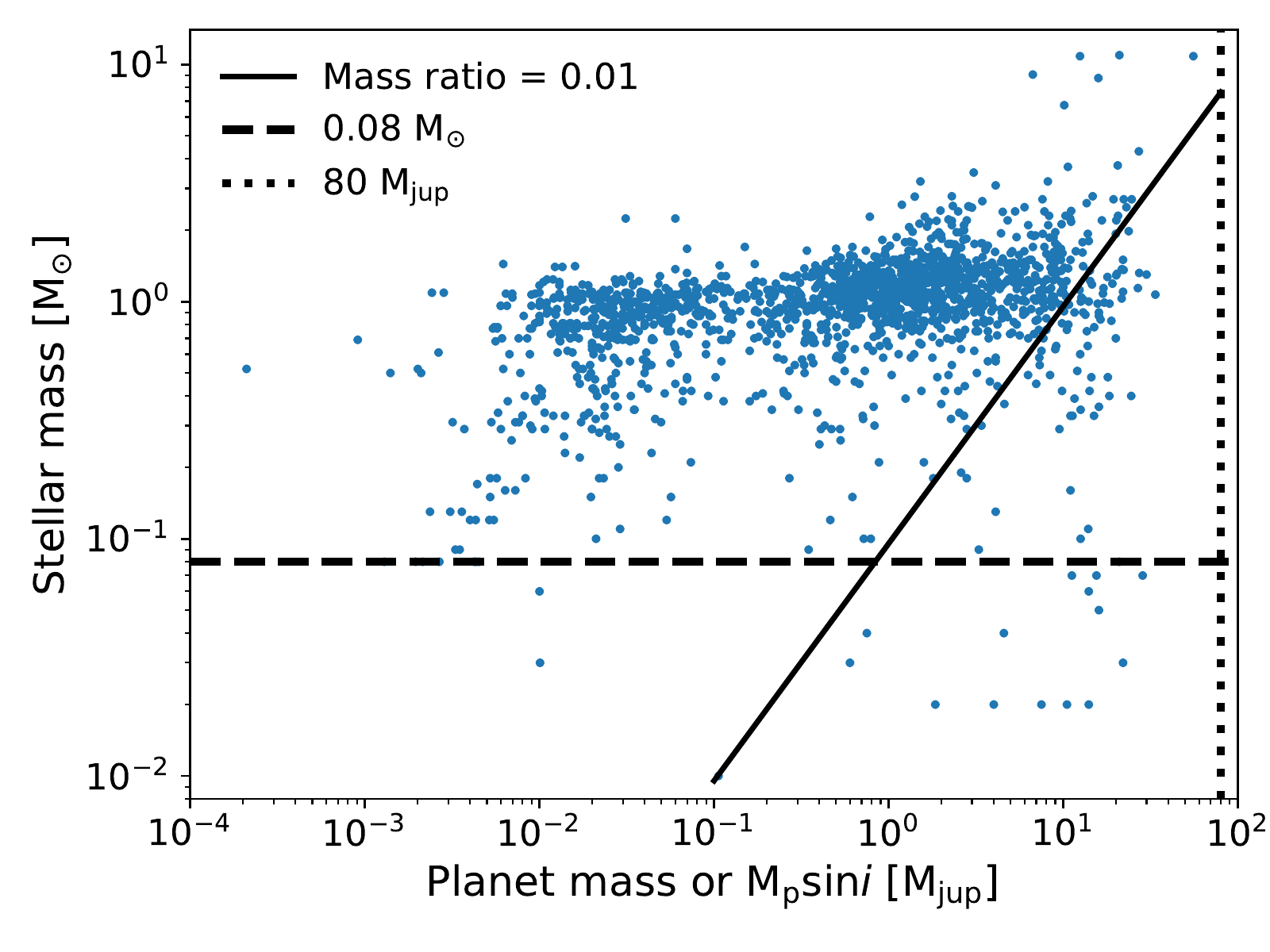}
    \caption{Currently measured planet masses  (or minimum mass) and those of the parent stars. Solid line depicts the planet-to-star mass ratio equal to 1\%. Planets at left of the solid line have mass ratios less than 1\% and correspond to $\sim$95\% of the total planet sample currently available in Exoplanet Archive. Dashed and dotted lines correspond to the mass limits to be considered stars and planets, 0.08 $\mathrm{M_{\odot}}$ and 80 $\mathrm{M_{jup}}$, respectively.}
    \label{masses_ratios}
\end{figure}

\rev{Here we describe the preparation of the data before running the simulations.} In the context of object-oriented programming, we define a Planetary System object by specifying the stellar mass and radius \revi{and the harbored planets}. Transit times fitting will be operated over Planetary Systems. 
Then, we define the planets by establishing the information about their transit times ephemeris and the allowed parameter space.
The transit ephemeris per planet should include the epoch number, time of transit and lower and upper timing uncertainties. 
\revi{In the simulations, the transit epochs (integer transit numbers) of all the planets are counted starting from 0 after $t_0$. Thus, the user must be aware that the epoch numbers are properly labeled and referenced to the same reference time $t_0$.}

\revi{The time span of the simulations is automatically chosen to encompass the full time of observations in the ephemeris data.}
If no time step is defined for the simulations, it is automatically set to be 30 steps per orbit of the internal planet ($\sim$3.33\% of the internal planet period). With this time step \cite{2014ApJ...787..132D} demonstrated the effectiveness in determining transit times with an accuracy within 10 seconds for a $\sim$22.3 days period planet. A time step < 5\% of the internal planet period is recommendable to reach that accuracy. 

Regarding the parameter space, the tool requires specific boundaries for each parameter in order to perform an effective sampling, avoiding non-physical regions or redundant information. Each planet in the Planetary System must define its own parameter space.
If there is not constraint for any or none of the parameters in $\Theta_j$, a set of default boundaries are established. 
By default, masses are restricted to be in the range from 1 Lunar mass to 80 Jupiter masses. However, when information of the stellar mass is supplied, the upper planetary mass limit is recalculated to be at most 1\% of the stellar mass. This is done to keep valid the N-body Hamiltonian internally solved by \texttt{TTVFast}. 
In Figure \ref{masses_ratios}, we show the currently measured planet masses $\mathrm{M_p}$ or $\mathrm{M_p\sin}i$ from Exoplanet Archive\footnote{\url{https://exoplanetarchive.ipac.caltech.edu}}. It is shown the planet mass limit corresponding to 1\% of the stellar mass.
We find that $\sim$95\% of the currently known measured planet masses have mass ratios with the parent star lower than 1\%, and thus the selected cutoff of 1\% is statistically valid for most of the currently known planets.
The default period space is limited to be between 0.1 and 1000 days. 
The eccentricity is limited between 1e-6 and 0.9, where the lower limit is different from exactly 0 to avoid undefined orbital angles. Inclination is defined between 0$^{\circ}$ and 180$^{\circ}$. 
\rev{In the case of periodic boundaries for the argument of periastron, $\omega$, and mean anomaly, $M$, fixing boundaries in the range 0$^{\circ}$ to 360$^{\circ}$ could lead to an improper sampling process when the solution is near to these borders. 
This problem is solved internally by parameterizing the angles by means of  $\omega$ + $M$ and $\omega$ - $M$. The boundaries in the parameterized space encompass 720$^{\circ}$. At the end of the sampling process, the results in the parameterized space are mapped to be between 0$^{\circ}$ and 360$^{\circ}$, for the individual $\omega$ and $M$. It is an internal process so the user only defines these angles between 0$^{\circ}$ and 360$^{\circ}$. 
Ascending node also must be defined between 0$^{\circ}$ and 360$^{\circ}$.}
These default ranges can be modified by the user to better constrain any parameter or to keep it as fixed. In the \revi{case of assuming constant parameters, these are} not part of the sampling process.

After setting the parameter space, \texttt{Nauyaca} normalize the boundaries of all the planets to dimensionless boundaries between 0 and 1. The whole sampling process (both for optimization and the MCMC) is performed with the normalized boundaries and returned to physical values at the end of the runs. This normalization is done to remove the differences in orders of magnitude between parameters, which enhances the sampling performance.

\subsubsection{Optimization module}\label{subsec_opt}  

\begin{figure}
    \hspace*{-0.5cm}
    \includegraphics[scale=0.48]{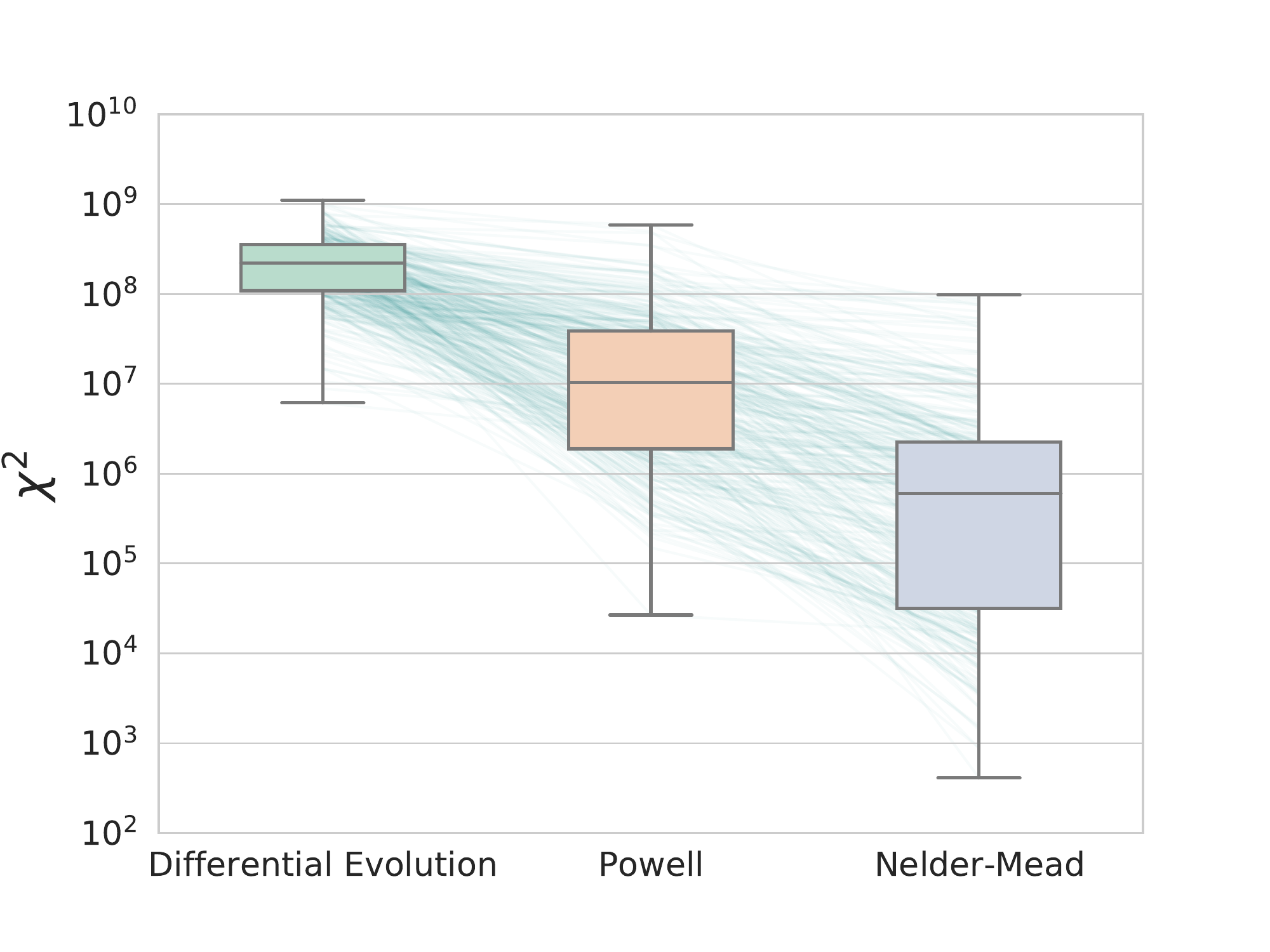}
    \caption{Example of the gradual $\chi^2$ minimization when running the sequence of  algorithms Differential Evolution, Powell and Nelder-Mead. \revi{Background lines connect the $\chi^2$ values of the solutions in the same run. Box edges show from bottom to top, the first and third quartiles of the data. Lines in the middle of the boxes denote the medians. Whiskers extend from the minimum to maximum values.}
    }
    \label{opt_sequence}
\end{figure}

It combines minimization algorithms ordered sequentially to reach solutions $\Theta_j$ that best explain the transit times. These results can be used as initial guesses for the MCMC. We tested many algorithms and their calling order and found that the sequence Differential Evolution \citep[DE; ][]{Storn:1997:DEN:596061.596146},  Powell \citep[PW; ][]{10.1093/comjnl/7.2.155} and Nealder-Mead \citep[NM; ][]{Gao2012} progressively minimize  the total $\chi^2$ (eq. \ref{eq_chi2}). 
This sequence is not arbitrary but reflects the nature of exploration of each algorithm.
First, DE is capable of exploring a large parameter space without the necessity of requiring an initial guess. The algorithm is fully stochastic and neither information about the smoothness of the space is required. The only mandatory information needed is the parameter space to explore, which have been set in the Setup module. A drawback of this algorithm is the slow convergence rate, \revi{and thus the outputs of this method could correspond to non-converged solutions.}
Even so, these solutions are \revi{better than any random proposal in the parameter space. Therefore, these are} used as an initial guess to run PW, which is suitable to perform a minimum searching, assuming that the space around the starting point is continuous although complex. 
Finally, NM takes the solution previously found by PW and performs a downhill simplex method assuming that locally the parameter space is smooth and unimodal.

\revi{We show in Figure \ref{opt_sequence} an example of the progressive $\chi^2$ minimization for the transit times fitting of a two-planet system with 190 transit times in total. For this example we performed 320 realizations. Background lines connect solutions of the same run, where  a gradual descending of the $\chi^2$ throughout the sequence is observed.
The box plots show the intervals of the $\chi^2$ achieved with the different algorithms, where typically the $\chi^2$ is reduced around 4-5 orders of magnitude between the first and last methods.
}

Performing multiple realizations of this process enhances the chance of finding a global minimum but also provides us with a global view of the most probable regions in the parameter space (this will be addressed in detail in subsection \ref{initialization_selection}). It is a fast way of finding a starting point region with valuable information that can help to delimit the searching radius where the MCMC routine will explore.

\subsubsection{MCMC module} 

We adapted a Markov \revi{chain} Monte Carlo method to explore the parameter space. 
We chose the Parallel-Tempering sampler, \texttt{ptemcee} \citep{PhysRevLett.57.2607,B509983H, Vousden_2015,Foreman_Mackey_2013} which is a well-suited sampler for exploring multi-modal and a highly dimensional parameter space, such as in the case of planetary systems, whose parameter space increases as $\sim$7 times the number of planets (one for mass and six orbital elements).
The main idea behind the technique is  using armies of walkers belonging to a "temperature ladder" that explore the parameter space in different detail. The posterior distribution $\pi$ is modified according to the temperature $T$, given by $\pi_T \propto L(\Theta)^{1/T} p(\Theta)$, being $L$ and $p$ the likelihood and the prior distributions, respectively. Walkers belonging to hotter temperatures sample more efficiently from the prior and those in colder temperatures better sample regions with high probability. Walkers in different temperatures have a probability of swapping positions in the parameter space that depends on their current positions and temperatures, such that those in colder temperatures can also explore from the prior and vice-versa. This technique avoids the walkers getting stuck in local solutions and explore more efficiently the whole parameter space where other standard samplers could fail \revi{(see \cite{Vousden_2015} for more details about this technique). }

\section{Mock Catalog} \label{sec:catalog}

\begin{figure*}
    \hspace*{-1.5cm}
    \includegraphics[width=1.15\textwidth]{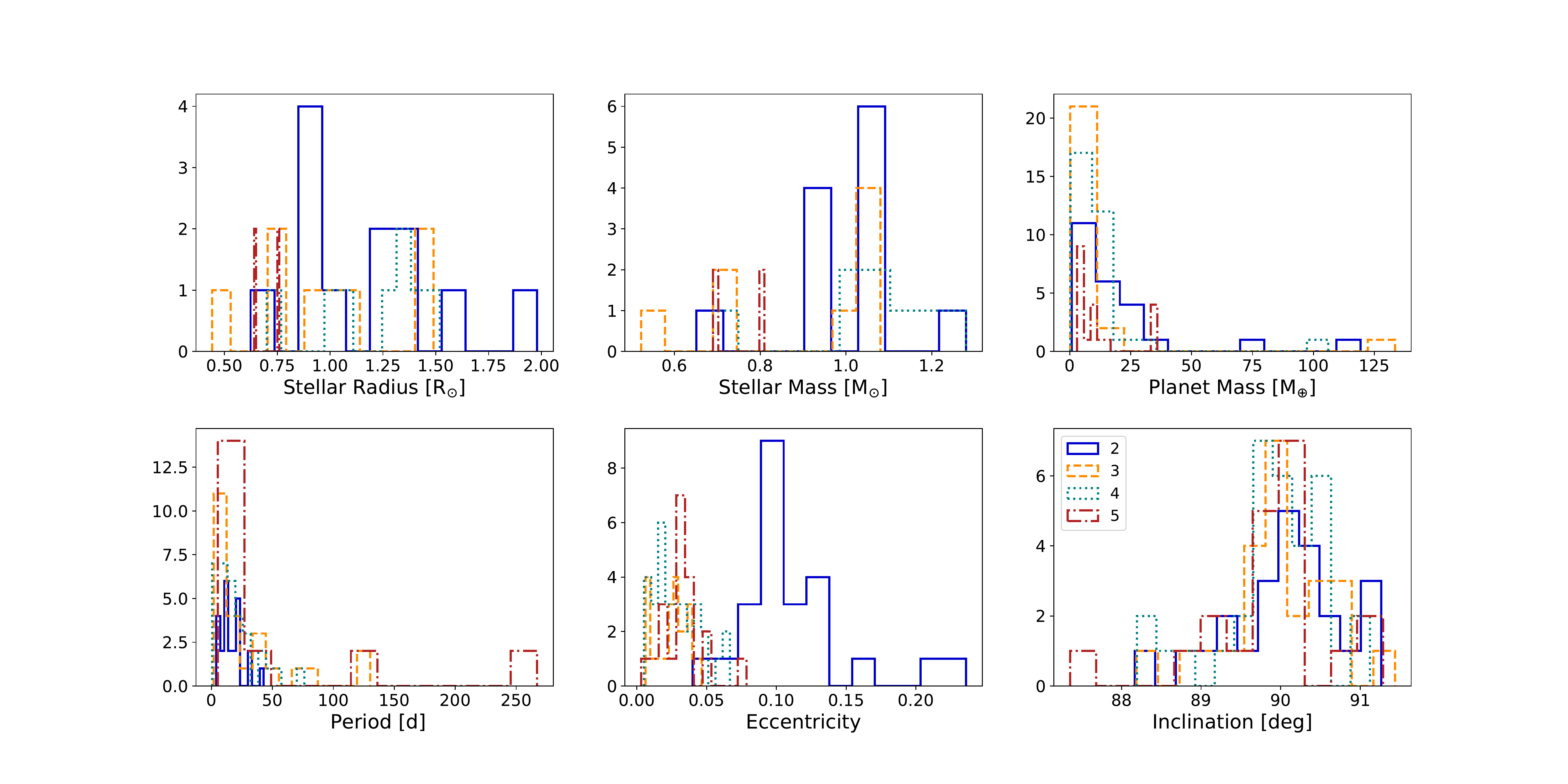}
    \caption{Histograms of the stellar and planetary parameters that compose the full mock catalog. Different colored lines represent the properties grouped by planet multiplicity for systems with two (solid blue), three (dashed orange), four (dotted cyan) and five (dash-dotted red) planets.
    Remaining orbital elements (not shown here) are taken from uniform distributions. Appendix \ref{catalog_tables} contains the data tables of these histograms.}
    \label{catalog_parameters}
\end{figure*}

We create a  catalog of synthetic planetary systems, defining stellar masses and radius, and planetary masses and orbital elements. We calculated transit times for these planets
and apply \texttt{Nauyaca} to this synthetic catalog in order to test the efficiency in recovering the planetary parameters (catalog entries) that gave rise to the synthetic transit times per planet. 
Throughout the text, we will refer as $trues$ to the parameter values reported in our mock catalog.

\subsection{Catalog creation}\label{catalog_creation}

In order to set up planetary systems as realistic as possible, we selected masses and radii for the stellar hosts; masses, radii and orbital periods for planets from the available data of confirmed planetary systems from the Exoplanet Archive\footnote{From available data at date January 24th, 2020.}. This was done to create synthetic planetary systems with stellar and planetary parameters based on observations.
The catalog includes different planetary multiplicities (number of planets in the system), ranging from two up to five planets. Below we describe the procedure followed to create the catalog.

First, we selected systems with planets discovered by the transit method. Second, we selected systems with reported stellar masses and radii. For those with unavailable data but with reported effective temperature, surface gravity and metallicity, we derived the stellar mass or radius from the work of \cite{2010A&ARv..18...67T}. 
Third, we selected those systems where all of their planets have reported masses, radii and periods (systems where planets have these parameters unreported were discarded).
This procedure reduces significantly the number of planets, dropping until 2.5\% (105 planets) of the original observed planets. 
In order to increase the number of synthetic planetary systems we applied an over-sampling technique \citep[SMOTE;][]{2011arXiv1106.1813C} which makes new samples by interpolation of the $k$-nearest neighbors. Thus, we triplicate the number of planetary systems and their parameters, namely, stellar mass and radius, and planetary masses, radii and periods.
The remaining five orbital elements to complete a unique planetary configuration, was made with random uniform distributions in the intervals: $ecc$ [0.05, 0.2], $inc$ [89.5, 90.5] deg, $\omega$ [0, 360] deg, $M$ [0, 360] deg, $\Omega$ [88, 92] deg. We restricted the intervals of the inclination $inc$ and ascending node $\Omega$ to get near coplanar and prograde orbits. These restrictions in the construction of the catalog are independent of the tool test. In practice, we can expand the boundaries of any parameter to be explored, as long as they have a physical meaning (see section \ref{planet_boundaries}).

Once the complete set of planetary parameters are established, an N-body integration is performed during $10^6$ orbits of the internal planet using REBOUND \citep{2012A&A...537A.128R, 2015MNRAS.452..376R}. 
We used the chaos indicator MEGNO \citep{2003PhyD..182..151C, 2011IJNLM..46...23M, 2015MNRAS.452..376R} to test the dynamical stability of the proposed planetary system. A MEGNO value around $\sim 2$ indicates quasi-periodic motion (regular stable orbits).
If the set of parameters of the proposed planetary system turns out to be stable (without planetary ejection or with 1.7 < MEGNO < 2.3) then we append the final state of the N-body simulation as an entry in the catalog. 
\rev{It should be pointed out that the orbital elements of these entries correspond to the osculating elements at the end of the stable N-body runs and they are not the same that the random values taken from the intervals of parameters indicated above. }
Finally, we use these entries as initial conditions to run TTVs simulations encompassing 130 transits of the internal planet using \texttt{TTVFast}. We selected that number to mimic the number of transits for planets with periods < 10 days, during $\sim 3$ years of observation. 
We use a fixed number of transits rather than a fixed time span since it is more relevant for the inversion problem \citep{2008ApJ...688..636N}.
As a result we get 130 transit times ephemeris for all the internal planets, and a few less for the remaining planets according to their periods and orbital configurations. 

We assign synthetic errors to these transit times according to the analytical approximation of the timing precision \citep{2018haex.bookE...7A}, 

\begin{equation}
\sigma_t = \frac{1}{ \sqrt{2}} \ \tau^{1/2} \dot{N}^{-1/2} \delta^{-1}
\label{sigma_tau}
\end{equation}

\rev{\noindent where $\tau$ is the approximate duration of the transit ingress $\tau \approx 2.2 \mathrm{min} \left(R_p/R_{\oplus}\right) \left(M_{*}/M_{\odot}\right)^{-1/3} \left(P/10d\right)^{1/3} $ which is a function of planetary radius $R_p$, orbital period $P$ and stellar mass $M_*$, 
$\delta= (R_p/R_*)^2$ is the depth of the transit, and $\dot{N}$ is related with the Poisson noise due to the count rate of the star. We assumed $\dot{N}$ to be constant and equal to 1$\times$10$^7$ $e^-$/min to mimic the value in the $Kepler$ CCDs for a star of magnitude $\sim$12 in the $Kepler$ band  \citep{2011ApJS..197....6G}.}
Uncertainties from Eq. \ref{sigma_tau} take planetary parameters derived from the $Kepler$ photometry, and therefore, our study is centered in the characterization of $Kepler$-like transits.
\revi{Finally, we added white noise to the transit time models assuming a normal distribution with the mean equal to the true transit times and with standard deviation equal to the typical uncertainties given by Eq. \ref{sigma_tau}. }

The full mock catalogue is composed by twelve two-planet systems, eight three-planet systems, eight four-planet systems and four five-planet systems, giving a total of 32 systems harboring 100 planets in total. \rev{We remark that  all the planets in the catalog transit, and thus our current study does not include the characterization of non-transiting planets.} Tables in Appendix \ref{catalog_tables} include these parameters of the simulated mock catalog. Figure \ref{catalog_parameters} shows the distributions of the parameters in the catalog according to the planetary multiplicity.

The mock catalog of planetary parameters produces a variety of TTVs with different properties, namely, amplitudes, periodicity and time span. TTVs signals in our catalog have amplitudes consistent with almost zero minutes, to amplitudes reaching 180 minutes, and  with an overall mean of $\sim$\revi{18} minutes. 
Mean TTVs amplitudes grouped by multiplicity reach \revi{21, 21, 10 and 23} minutes for two, three, four and five-planet systems, respectively.
It translates into a wide range of signal-to-noise ratios, ranging from \revi{$\sim$1 up to $\sim$400} with a mean of $\sim$23 \citep[using the definition of the ratio of the TTVs amplitude and the timing uncertainty; ][]{NESVORNY2019101507}.
There is also a variety in observation time span, ranging from $\sim$230 to $\sim$5800 days with a mean of 1200 days.

\section{Settings}\label{sec:settings}

\begin{figure*}[t!]
    \includegraphics[scale=0.45]{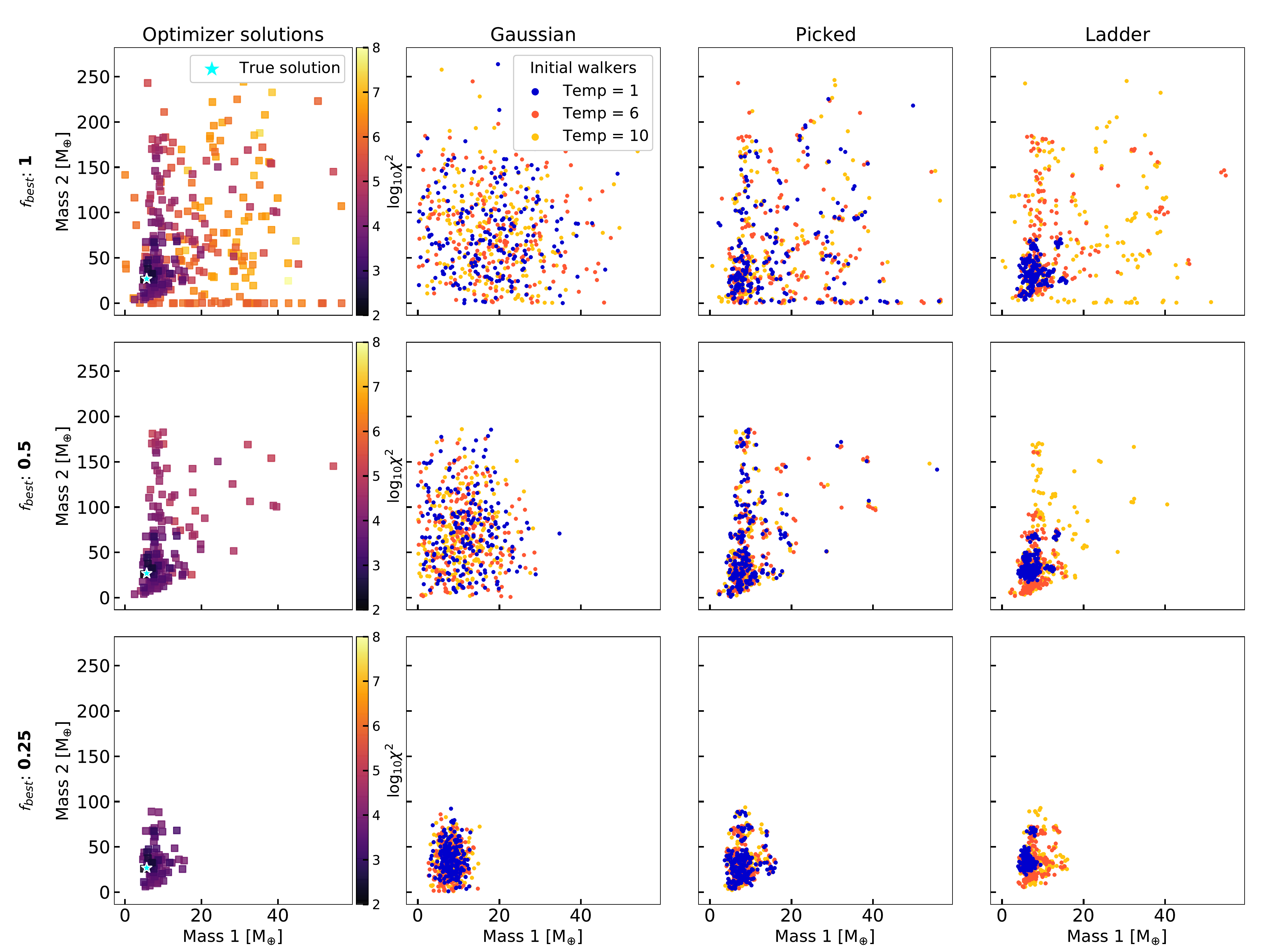}
    \caption{
    \revi{
    Comparison between three strategies to initialize walkers for the MCMC with the information from the optimizers. This example considers the walkers initialization over the two-dimensional space of the masses, for a system with two planets (\revii{ID=pl2\_id52}; Table \ref{catalog2p}). 
    The first column of panels shows the solutions found by the optimizers.
    The position of the $true$ masses of these planets is shown by the cyan star. Color bar shows the logarithm of the $\chi^2$ for these solutions. 
    Remaining columns show the walkers initialization using different strategies (see text for a full description). Colored dots are the initial walkers belonging to the coldest (1; blue), intermediate (6; orange) and the hottest (10; yellow) temperatures.
    For clarity, walkers belonging to intermediate temperatures are not shown.
    Rows from top to bottom show the parameter $f_{best}$ taking the 100\%, 50\% and 25\% of the best solutions from optimizers, from which the initial walkers are drawn.
    }
    }
    \label{initialization}
\end{figure*}

In this section we outline the election of the parameters to run the tool, including the \revi{restrictions of the parameter space, the election of fine-tuning parameters for the MCMC and the usage of the optimizer results to determine a reasonable starting point for the sampling process. }

\subsection{Planetary boundaries}\label{planet_boundaries}

As described in section \ref{subsec:setup}, \texttt{Nauyaca} requires specific boundaries for each planetary parameter in $\Theta$. These boundaries delimit the parameter space that will be explored. Here, we will discuss how to delimit the parameter space \revi{(including the establishment of constant parameters) before carrying out the recovery test over the mock catalog.}

From fitting real transit observations, is possible to determine the planetary radius ratio, the orbital period from consecutive transits and the impact parameter. In the case of the last two, we measure from current data of observed exoplanets (from Exoplanet Archive) that orbital periods are determined, on average, with uncertainties $\sim10^{-4}$ days. On the other hand, orbital inclinations are on average, better determined than 1$^{\circ}$ (also according to Exoplanet Archive).
Considering nearly coplanar orbits, imply that line of nodes are nearly aligned and mutual inclinations are close to 0$^{\circ}$, which can help to reduce the dimensionality of the problem.
Observed transiting exoplanets have shown to have small mutual inclinations, typically  lying below 3$^\circ$ \citep[][]{2012ApJ...761...92F,2014ApJ...790..146F} which have been demonstrated to have a negligible effects on TTVs \citep{2009ApJ...701.1116N,2014ApJ...790...58N, 2016ApJ...828...44H}. Furthermore, because the orbital inclination is a well-restricted observational parameter (with an error mean of 0.78$^{\circ}$ with a dispersion of 2.2$^{\circ}$ according to data from Exoplanet Archive), it can be kept as a fixed parameter. 
Additionally, in test runs we let the inclination vary between $\sim80^{\circ}-100^{\circ}$ (which is a comprehensible range of inclinations that allows the transits) and confirm that this angle has limited effects on the results for the transit time models.
Nearly aligned ascending nodes represent prograde orbits and anti-aligned represent retrograde orbits. Thus, in order to model nearly coplanar orbits we keep fixed the orbital inclinations to their $true$ values, $inc_{j,True}$. 
We also keep fixed the ascending node of the internal planets $\Omega_{1,True}$ ($\approx$ 90$^o$), which by construction of the coordinate system can be fixed.

For our recovery test, we delimited the planetary parameter space to these ranges, bracketing the lower and upper limits: mass [0.0123, 10 mass$_{j,True}$] $\mathrm{M_{\oplus}}$, $P$ [$P_{True} - \delta t$, $P_{True} + \delta t$] days, $ecc$ [$10^{-5}$, 0.3], $inc$ (fixed) [$inc_{True}$] deg, $\omega$ [0, 360] deg, $M$ [0, 360] deg  and $\Omega$ [70, 110] deg. 
Here, the lower and upper boundaries for $masses$ correspond to approximately a Moon mass and 10 times the $true$ planetary masses, respectively.
The boundaries in period $P$ are around the $true$ period of the planet with a width $\delta t$ corresponding to a typical observed period error given the orbital period itself. We estimated
$\delta t$  by doing a linear regression between observed period errors as a function of the orbital period (data from Exoplanet Archive), such that $\delta t = m P + b$ [days], where we determined $m= 2.11\times 10^{-5}$ and $b=4.4019\times 10^{-5}$.
The lower limit in eccentricity ($ecc$) was established small but different from 0 in order to avoid undefined argument of periastron, whilst the upper limit was allowed to get values up to 0.3. This chosen range of eccentricities is compatible with the 80\% of the currently observed planet eccentricities. 
Argument of periastron ($\omega$) and mean anomaly ($M$) take the usual definition between 0$^o$ and 360$^o$.
The boundaries of the ascending nodes ($\Omega$) for all the planets, except the internals, were restricted to a search radius around $\approx \Omega_{1,True} \pm 20$ deg, and thus considering only prograde orbits for simplicity.

\subsection{Parameters for the MCMC}

We inspected the dependence of the parameters that govern the MCMC.
The chosen parallel-tempering MCMC method \citep{Vousden_2015} is fine-tuned by two main parameters, namely, number of temperatures (N$_{\mathrm{temps}}$) and maximum temperature  ($T_{\mathrm{max}}$) \revi{to build the temperature ladder.} 
\revi{The sampling performance} also depends on the number of walkers per temperature (N$_{\mathrm{w}}$) and finally on the number of iterations (N$_{\mathrm{iter}}$) per chain.
We carried out many tests with combinations of these parameters and find that a \revi{N$_{\mathrm{temps}}\lesssim$ 15 with N$_{\mathrm{w}}\lesssim 150$} in a run over N$_{\mathrm{iter}}\sim 5\times 10^5$ are enough in most cases to recover the $true$ parameters $\Theta_{j,True}$ from the mock catalog within 1-$\sigma$. 
We also note that a $T_{\mathrm{max}}\sim 10^2-10^3$ is a good election of the maximum temperature.
A $T_{\mathrm{max}}=\infty$ as suggested in \citet{Vousden_2015} is not adequate for our purposes since walkers belonging to this temperature would propose steps outside our predefined boundaries. We confirm this behavior on several occasions during our tests.
Other parameters to control the dynamics of the temperature adjustment are internally defined within \texttt{Nauyaca} following the suggestions by \citet{Vousden_2015}.

\revi{For the recovery test presented in this work, we impose uniform log-priors for simplicity with the functional form,

\begin{equation}
\log (\mathcal{P}_k) =\left\{\begin{matrix}
0 &  b_{\mathrm{low}} \leq k\leq b_{\mathrm{upp}} \\
-\infty  & \mathrm{otherwise }
\end{matrix}\right.,
\label{eq_logprior}
\end{equation}

\noindent where $b_{\mathrm{low}}$ and  $b_{\mathrm{upp}}$ correspond to the lower and upper boundaries for the $k$-th planetary parameter. 
\revii{Table \ref{table_priors} shows a  list of the model parameters for each planet as well as the adopted range of uniform priors.} The election of these validity ranges  has been described previously in subsection \ref{planet_boundaries}.  
\revii{From these parameters, $inc$ was set as a constant to the $true$ inclination, $inc_{True}$, taken from the mock catalog. For each system,  $\Omega$ consider uniform priors except for the innermost planet (Planet 1 in the catalog entries) which is set to the true $value$ $\Omega_{1, True}$.}
Thereby, the posterior probability (up to a constant) is given by the sum of the log-likelihood (Eq. \ref{Loglikelihood}) and the log-prior (\ref{eq_logprior}) functions.
In practice, any other prior functions can be supplied to \texttt{Nauyaca} to calculate the posterior probability.
}

\begin{table}[t!]
\caption{Model parameters and the selected priors for the transit timing fit. $\mathcal{U}$[$b_{\mathrm{low}}$, $b_{\mathrm{upp}}$] denotes the uniform ranges between lower and upper boundaries. Single values are the fixed parameters in the simulations. Parameters with label $True$ take the data values from the mock catalog (see Appendix \ref{catalog_tables}). See text for details.
}
\label{table_priors}
\begingroup
\setlength{\tabcolsep}{14pt} 
\renewcommand{\arraystretch}{1.2} 
\begin{tabular*}{\columnwidth}{clc}
\hline \hline
\multicolumn{1}{l}{Parameter} & \multicolumn{1}{c}{Prior}  & \multicolumn{1}{c}{Units} \\
\hline 
$mass$    &  $\mathcal{U}$[0.0123,\  10 $\times$  mass$_{True}$ ]   & M$_{\oplus}$ \\
$P$      &   $\mathcal{U}$[$P_{True}- \delta t$,\   $P_{True}+ \delta t$ ] & days  \\
$ecc$    &   $\mathcal{U}$[1e-06,\  0.3] \\
$inc$    &   $inc_{True}$  &  deg \\
$\omega$ &   $\mathcal{U}$[0,\  360]  &  deg\\
$M$      &   $\mathcal{U}$[0,\  360]  & deg  \\
$\Omega$ &   $\mathcal{U}$[70,\  110]\ or\ $\Omega_{1, True}$   & deg \\
\hline
\end{tabular*}
\endgroup
\end{table}

\subsection{\revi{MCMC initialization from Optimizer results}}\label{initialization_selection}

Solving the inversion problem of a number of interacting planets, N$_{\mathrm{pla}}$, implies to explore a parameter space of dimensions $\sim7$N$_{\mathrm{pla}}$. The nature of the problem is computationally demanding since an N-body integration should be done at each iteration per walker. The wall clock time also increases with the observational time span. Thereby, choosing a strategical initial guess for walkers adapted for the parallel-tempering MCMC is crucial to minimize these side effects.

\revi{Starting random points from the prior function would be a reasonable choice assuming that we have an informed previous knowledge about the parameters. In general, it would not be the case for planets characterized for the first time. 
This motivate us to use optimizer results to make an educated initial guess about the planetary parameters. 
Optimizers} take advantage of both, a low computing time consumption in comparison with a full MCMC run \revi{(bewteen 1 and 3\% of the total time)}, \rev{but also, in}  finding many modes in the parameter space since realizations are independent among themselves. Although these solutions could be just rough approximations of more detailed solutions, \revi{they could contribute to identify high probability regions suitable for initialization. Initializing walkers near an optimum sensible place is better than any  random point in the parameter space \citep{2018ApJS..236...11H}. 
Even more, initialization using multi-start local optimization results is found to enhance the exploration quality and be more suitable for multi-chain methods \citep[\eg][]{Hug2013, Ballnus2017}.

We present three strategies implemented in \texttt{Nauyaca} to initialize walkers using the information from the optimizers, namely, $gaussian$, $picked$ and $ladder$. 
In order to set initial values from these strategies, we first make a filter over  the total number of optimizer solutions (N$_{\mathrm{opt}}$) by sorting them according to their $\chi^2$. Then, we take a fraction ($f_{best}$; defined between 0 and 1) of that ordered list which includes the uppermost solution. That subset of solutions $\Theta_{opt}$ is then used to initialize walkers.

In Figure \ref{initialization}, we show examples of initialization using these strategies for many values of $f_{best}$.}
\revi{For the current example, we focus on the  masses space of two planets identified with \revii{ID=pl2\_id52} in Table \ref{catalog2p}.} We performed 320 realizations of the optimizers to draw  \revi{an initial walker population} for a ladder of $\mathrm{N_{temps}}=10$ temperatures, considering different values of the parameter  $f_{best}$. 
\revi{The individual solutions from  $\Theta_{opt}$ are shown with squares in the first column, and the cyan star shows the position of the $true$ masses.
Colored dots are the initial walkers belonging to different temperatures. As $f_{best}$ is reduced, solutions with high $\chi^2$ are discarded. 
We detail these initialization strategies adapted to the parallel-tempering MCMC:}

\begin{enumerate}
    \item $Gaussian.$ Walkers are drawn below a gaussian centered in the mean of $\Theta_{opt}$ with a $1-\sigma$ value corresponding to the data dispersion, for each dimension. 
    \revi{This is the simplest initialization method and possibly the most frequently used. The main difference is that here the mean and standard deviation are based on the independent random realizations and not on previous knowledge of the planetary parameters (for example, masses from radial velocity measurements).}
    Thus, if the optimizers find a well-restricted solution for any dimension, the MCMC will be able to find the global high probability region faster in contrast with an $uniform$ \revi{random initialization}. 
    From Figure \ref{initialization}, it is seen that using this \revi{strategy} while reducing $f_{best}$ could \revi{help to identify the zone in the parameter space where the global minimum could exist.} 
    \revi{Thus, using this strategy most of the local modes are covered but there is not special attention in the individual modes or in the possible correlations between parameters.  }

    \item $Picked.$ Solutions from $\Theta_{opt}$ are randomly picked and the initial walkers are drawn in the vicinity of those solutions. If the optimizers find many modes, this \revi{strategy} ensures that walkers will be drawn around all the modes which could correspond to local minima \revi{or the global minimum}. 
    \rev{From Figure \ref{initialization}, it is seen that this strategy} confines the initial population of walkers as $f_{best}$ is reduced 
    \revi{while keeping the dependency between parameters.
    Here, walkers are equally distributed around any mode independently of their temperature and therefore all the modes are initially sampled with the same frequency.
    This could be suitable for solving problems where apparently there is not a unique region  where optimizer results agglomerate.
    }

    \item $Ladder.$ Solutions from $\Theta_{opt}$ are divided in an integer number of chunks equal to the number of temperatures. Walkers belonging to temperature 1 (the main temperature; blue \revi{dots}) are drawn from the first chunk, which includes the \revi{uppermost solution} (\ie with the lower $\chi^2$). Walkers belonging to the second temperature are drawn from the first and second chunks. 
    The same rule is followed for the rest of the temperatures \revi{until finally, walkers for the hottest} temperature (\revi{yellow dots) are drawn around} all the solutions in $\Theta_{opt}$. 
    \revi{From Figure \ref{initialization} it is seen that using this strategy, the modes with the lower $\chi^2$ are highlighted} as $f_{best}$ is reduced.
    Unlike the $picked$ \revi{strategy, $ladder$ assigns the  outstanding modes to colder walkers while hotter walkers sample the more disperse ones.} 
    \rev{It allows the exploration of other modes but avoiding getting stuck in these local minima. }

\end{enumerate}

The election of the best strategy and parameter $f_{best}$ can vary according to the problem. Ideally, a visual inspection of the \rev{optimizer solutions (as in Figure \ref{initialization}) could help to identify modes in the parameter space that allows to decide} how to initialize walkers.
In practice, a statistical indicator (\eg standard deviation) or a clustering method could be helpful to choose the \revi{initialization}: $gaussian$ for high dispersed data; $picked$ for highly multi-modal parameter spaces and $ladder$ for multi-modal with a main mode (as in the example of Figure \ref{initialization}). 

\revi{Note however that the usage of the optimization module and the proposed initialization strategies are not a mandatory step in \texttt{Nauyaca} previous to the implementation of the MCMC. 
Any initial walker population can be provided by the user as long as these proposals are inside the physical boundaries described in section \ref{subsec:setup}. Nonetheless, we empirically find that following this heuristic procedure notably enhances the MCMC performance at the expense of low computational cost. }

\section{Results and discussion} \label{sec:Rec_results}

\begin{deluxetable}{ ccccccc }
\tablecaption{Used parameters for the recovery test, according  to planet multiplicity. \label{table_parameters}
}
\tablehead{
\colhead{N$_{\mathrm{pla}}$} & \colhead{N$_{\mathrm{opt}}$}  & \colhead{N$_{\mathrm{Temps}}$} & \colhead{$f_{\mathrm{best}}$ (\%)} & \colhead{Walkers} & \colhead{\rev{Steps $\times 10^5$}} & \colhead{\rev{Thinning}}
}
\startdata
2            & 320        & 10      &   6.5     & \revi{80}      & 2.5  & \revi{100}\\
3            & \revi{416}        & 10      &   \revi{6.0}     & \revi{80}      & 3.5  & 100\\
4            & 512        & 10      &   5.0     & \revi{100}      & \revi{4.5}  & \revi{200}\\
5            & 640        & 12      &   7.5     & \revi{120}      & \revi{6.5}    & \revi{200}\\
\enddata
\tablecomments{Walkers refer to number of walkers drawn per temperature.}
\end{deluxetable}

We applied \texttt{Nauyaca} and the same fitting procedure to the mock catalog with the aim of inverting the process going from the synthetic transit times to the planetary parameters and then comparing with the original values in the mock catalog, which we will refer to as the $true$ values.
For each run, we kept fixed the inclinations and the ascending node of the first planet, as described in section \ref{planet_boundaries}. All the solutions are determined at the synthetic \revi{reference epoch $t_0$ = 0 days,} to match with the reference time of the catalog construction.

The procedure consists of the following steps: 1) Providing the $true$ stellar mass and radius, transit ephemeris per planet (mid-transit times) and initializing the parameter space; 2) running the optimizers and choosing the best $\sim 5\%-10\%$ of the solutions to initialize walkers using the $ladder$ \revi{strategy}, and 3) taking the data from step 2 as initial walkers population for running the MCMC over a fixed number of \rev{steps} and using a Gelman-Rubin statistic \citep[< 1.01;][]{GelmanRubin1992}  and a Geweke test \citep[Z-score < 1;][]{geweke1991evaluating}  to assess the convergence and stationarity of the chains.

We performed the \rev{same} procedure with the parameters summarized in Table \ref{table_parameters} according to the number of planets in the system (N$_{\mathrm{pla}}$). 
Since the dimensionality of the parameter space scales with the number of planets, we increase the number of optimizer realizations ($\mathrm{N_{opt}}$) and MCMC \rev{steps}, accordingly. 
\rev{Along the MCMC runs we did a thinning by saving the current state of the chains at predefined number of steps (shown in Table \ref{table_parameters}), which also allows us to diminish the memory requirements. 
At the end of the runs we measure the mean autocorrelation time of the averaged chains and we determined typical values between \revi{30} and \revi{90}, with a mean of 70 steps. With these values we determined the effective sample size, getting typical values between  \revi{2,400} and \revi{6,800} with a mean of \revi{3,800} independent samples.} \rev{For a pair of systems with 3 (\revii{ID=pl3\_id4}) and 4 (\revii{ID=pl4\_id3}) planets, we repeated the MCMC process with the same parameters in Table \ref{table_parameters} but changing the \revi{initialization} strategy from $ladder$ to $gaussian$. By doing this, we initialize the MCMC with less informative points and we find consistent results within 1-$\sigma$ with the initial results.}

Using 16 cores per job (\ie per planetary system), \texttt{Nauyaca} was able to fit two-planet systems in \revi{$\sim$15 hours}, on average. 
The time increased with increased complexity of the planetary system, reaching up to \revi{$\sim$5.6 days} for five-planet systems. 
Most of the time is spent on running the MCMC, since in comparison, the optimizers are quite fast, running  N$\mathrm{_{opt}}$ solutions (specified in Table \ref{table_parameters}) within \revi{10} minutes and \revi{2.5} hours depending on number of \revi{optimizer runs}.
Note however that the wall-clock time depends on many factors, as for example the \revi{number of planets,} the time span of observations and number of transits to fit, which in our case exceed $\sim$150 transits for two-planet systems (see number of transits per planet in Appendix \ref{catalog_tables}). 
Thus, lower computational requirements would be enough for simpler systems than those considered in this work.
All the simulations were performed with the supercomputer Miztli at the Universidad Nacional Aut\'onoma de M\'exico.

\begin{figure*}[ht!]
    \centering
    \includegraphics[scale=0.32]{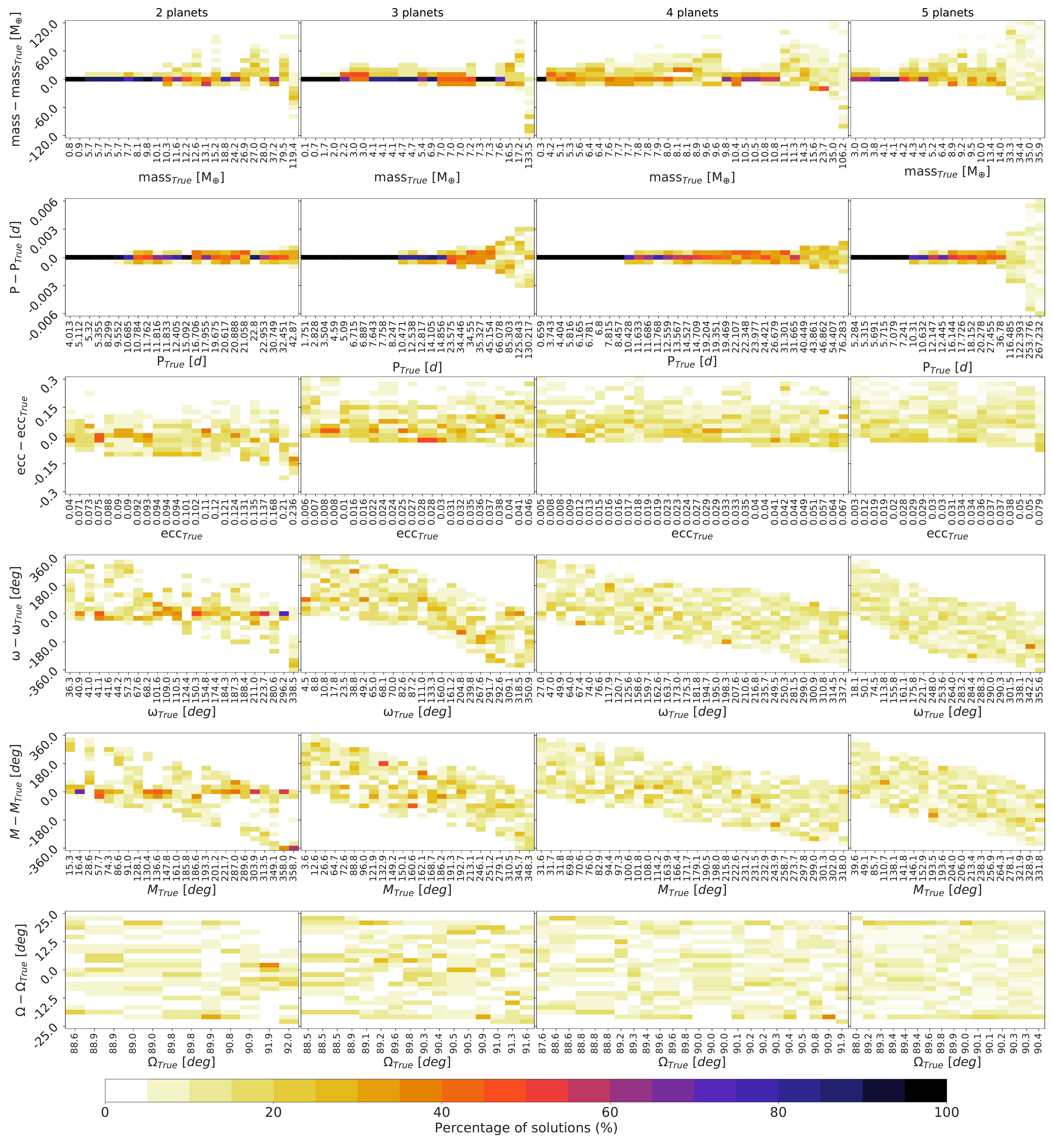}
    \caption{
    \rev{Optimizer results shown as the differences between the solutions found and the $true$ values}, grouped by the number of planets (columns) and for each planetary parameter (rows).  
    \rev{Individual sub-panels are binned along the horizontal axis by the $true$ planetary parameters. Note that these labels are arranged in increasing order but they are not equally spaced.}
    Color code shows the percentage of the solutions found by optimizers falling inside each bin. \rev{The total number of solutions for two, three, four and five-planet systems are \revi{21, 25, 26} and 48, respectively, which correspond to N$_{\mathrm{opt}}$ times $f_{\mathrm{best}}$, reported in Table \ref{table_parameters}.}
    }
    \label{opt_results}
\end{figure*}

\begin{figure*}[ht!]
    \centering
    \includegraphics[scale=0.32]{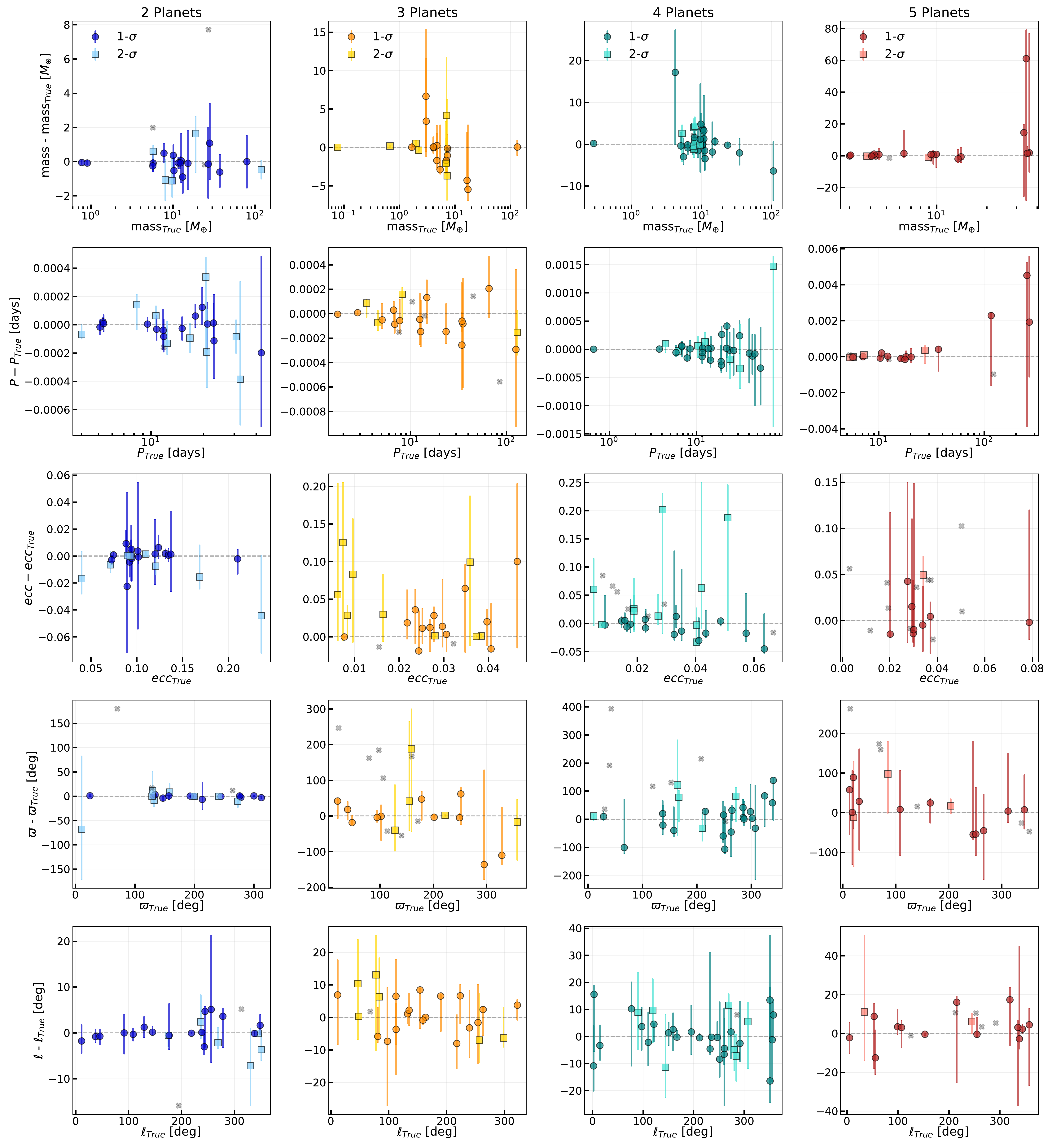} 
    \caption{\rev{Solutions found by the parallel-tempering MCMC compared with the $true$ planetary parameters.}
    The comparison is made, \rev{from top to bottom},  over the following planetary parameters: mass, period \revi{($P$)}, eccentricity \revi{($ecc$)}, periastron longitude ($\varpi$) and true longitude ($\ell$).  The recovery test was done over the transit times originated by these planets. Columns show the results according to the number of planets in each system. Horizontal axes denote the input data ($true$ planetary parameters) and vertical axes denote the \rev{differences between the posterior medians recovered by the MCMC and the $true$ parameters. }
    Dark color circles show data consistent within 1-$\sigma$ and light color squares those consistent within 2-$\sigma$. Vertical error bars are colored with dark or light colors for 1-$\sigma$ and 2-$\sigma$, respectively. Gray \revi{crosses denote the planets whose parameters were} inconsistent with the $true$ input data.  The percentage of planets which are consistent with the input data are summarized in Table \ref{table_percentages}.}
    \label{recovered}
\end{figure*}

\subsection{Optimizer results\label{sec_optimizers}}

\rev{We run the optimizers \revi{to explore} the parameter space with the aim of minimizing the differences between the observed and modeled transit times (Eq.  \ref{eq_chi2}). Since these} runs are independent from each other, increasing the number of realizations enhances the chance of finding more minima, that could correspond to local minima or the global minimum. 
We take a fraction \rev{(namely $f_{best}$) of the whole set of solutions with the best} $\chi^2$  to build a subset of solutions $\Theta_{opt}$, which will be used to initialize walkers.

Setting an optimum composition of $\Theta_{opt}$ is an interplay between N$_{\mathrm{opt}}$, $f_{best}$ and the number of temperatures. Thus, there is not a unique way of defining these parameters.
Though, we noticed that in most cases $f_{best}$ < 10\% is an appropriate fraction using N$_{\mathrm{opt}} \sim300-500$. 
We used $f_{best}$ shown in Table \ref{table_parameters}, according to planet multiplicity. That selection translates in \revi{21, 25, 26} and 48 solutions that compose $\Theta_{opt}$ for systems with two, three, four and five planets, respectively. 
Taking higher values of $f_{best}$ means taking more solutions from optimizers with increasing $\chi^2$. 
\revi{Usually solutions with high $\chi^2$ are located far from the meaningful modes, resembling to random proposals (as can be noticed from Figure \ref{initialization}). Thus, selecting almost all the optimizer solutions ($f_{best} \sim$ 1) could reduce the contribution of the optimizers to locate optimal initial regions to draw walkers.}

In Figure \ref{opt_results}, we show the results of the optimizers by measuring the \rev{differences between the solutions found} and the $true$ values. 
\rev{The solutions considered in this figure are part of the subset denoted previously as $\Theta_{opt}$. Therefore, most of the solutions from the optimizers with higher $\chi^2$ values were discarded.}
Color map indicates the \rev{percentage of solutions falling inside each bin.}
Darkest bins near to zero depict planets where $\gtrsim 50\%$ of the solutions better approximate to the $true$ solutions. \rev{Sub-panels with almost uniform yellow/white colors are those parameters less constrained by the optimizers.}

For the case of the planet masses,  \revi{90\%} of the solutions in $\Theta_{opt}$ are typically determined within a factor $\sim$\revi{2--6} with respect to the $true$ masses.
Even though the dispersion tends to increase with the increasing of the mass and the number of planets, the solutions approach to the correct mass with a relative low dispersion.
In the case of the periods, \revi{the solutions tend to span over all the allowed space.
We find this behavior is due to the narrowness between our selected lower and upper boundaries, which scales according to the period (see section \ref{planet_boundaries}). 
We noticed that when the boundaries are enlarged (for example, with a width of $\pm 0.01$ days) the solutions agglomerate around the $true$ periods.}
Eccentricities are in general well-restricted only for two-planet systems, where the dispersion reaches up  to \revi{$\sim$0.06} around the true values. 
For higher multiplicities the solutions scatter out, with a typical dispersion of \revi{$\sim$0.08.} We find that optimizers tend to overestimate eccentricities as the number of planets increases. 
For the argument of periastron ($\omega$) and mean anomaly ($M$), a similar behavior is found. \rev{They are better restricted only for two-planet systems, since for higher multiplicities the dispersion increases, tending to be evenly distributed over the parameter space.}
Ascending nodes are the less restricted angles given our established boundaries for this parameter. Note however that for all the recovery test we just considered limited prograde solutions (as described in section \ref{planet_boundaries}).

\revii{Although the initial search region explored by the MCMC would be narrowed down as a result of the optimization process, the chains are  initialized in these high density regions and explore all the allowed parameter space since the adopted uniform priors are not modified.}

\subsection{MCMC results\label{sec_mcmc}}

\begin{table*}[ht!]
\caption{Percentage of planets consistent with the $true$ parameters within 68\% (1$-\sigma$) and 95\% (2$-\sigma$) \revi{credible} interval. Percentages are given for specific planet multiplicities and for the full catalog.} \label{table_percentages}
\hspace*{-1cm}
\vspace*{0.5cm}
\resizebox{\textwidth}{!}{
\begin{tabular}{ccccccccccc}
\hline \hline
\multicolumn{1}{c}{\multirow{2}{*}{Planet multiplicity}} & \multicolumn{2}{c}{Mass}              & \multicolumn{2}{c}{Period}                   & \multicolumn{2}{c}{Eccentricity}             & \multicolumn{2}{c}{Periastron longitude ($\varpi$)} & \multicolumn{2}{c}{True longitude ($\ell$)} \\ \cline{2-11} 
\multicolumn{1}{c}{}                                     & 1-$\sigma$ & \multicolumn{1}{c|}{2-$\sigma$} & 1-$\sigma$ & \multicolumn{1}{c|}{2-$\sigma$} & 1-$\sigma$ & \multicolumn{1}{c|}{2-$\sigma$} & 1-$\sigma$    & \multicolumn{1}{c|}{2-$\sigma$}   & 1-$\sigma$               & 2-$\sigma$               \\
                                                         & (\%)       & \multicolumn{1}{c|}{(\%)}       & (\%)       & \multicolumn{1}{c|}{(\%)}       & (\%)       & \multicolumn{1}{c|}{(\%)}       & (\%)          & \multicolumn{1}{c|}{(\%)}         & (\%)                     & (\%)                     \\ \hline
2 (24 planets)                       & 66         & \multicolumn{1}{c|}{87}         & 58         & \multicolumn{1}{c|}{95}         & 66         & \multicolumn{1}{c|}{100}        & 54             & \multicolumn{1}{c|}{87}            & 66                   & 91                   \\
3 (24 planets)                       & 58         & \multicolumn{1}{c|}{91}         & 62         & \multicolumn{1}{c|}{80}         & 54         & \multicolumn{1}{c|}{91}         & 45             & \multicolumn{1}{c|}{66}            & 70                   & 95                   \\
4 (32 planets)                       & 75         & \multicolumn{1}{c|}{96}         & 78         & \multicolumn{1}{c|}{100}        & 47         & \multicolumn{1}{c|}{78}         & 62             & \multicolumn{1}{c|}{78}            & 75                   & 96                   \\
5 (20 planets)                       & 85         & \multicolumn{1}{c|}{95}         & 65         & \multicolumn{1}{c|}{85}         & 40         & \multicolumn{1}{c|}{45}         & 55             & \multicolumn{1}{c|}{70}            & 65                   & 75                   \\
\hline
Full catalog (100 planets)           & 71         & \multicolumn{1}{c|}{93}         & 67         & \multicolumn{1}{c|}{91}         & 52         & \multicolumn{1}{c|}{80}         & 55             & \multicolumn{1}{c|}{76}            & 70                   & 91                   \\ \hline \hline
\end{tabular}}
\end{table*}

\begin{figure}[t!]
    \includegraphics[scale=0.35]{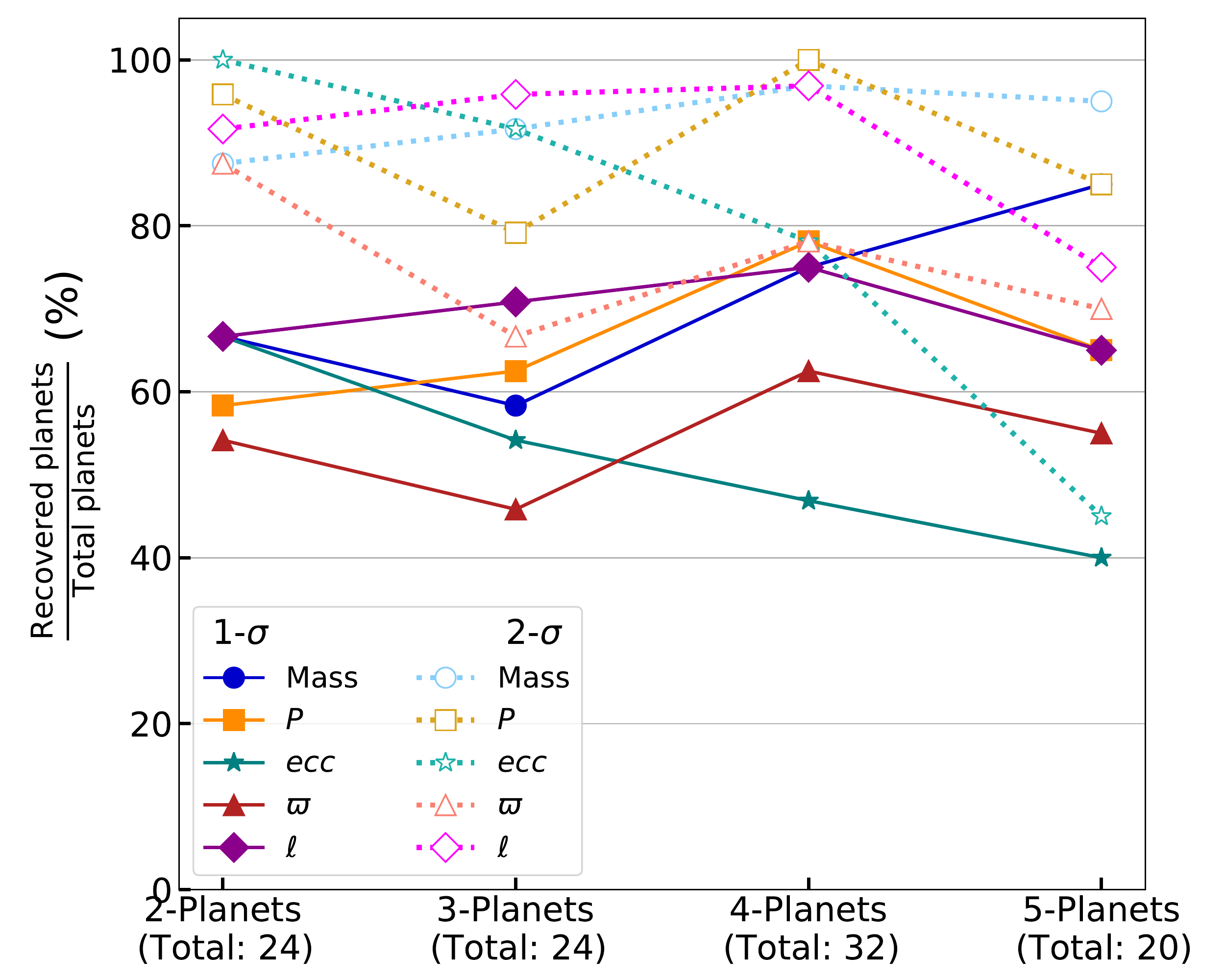}
    \caption{\rev{Percentage of planets whose results from the MCMC are consistent with the input parameters in the mock catalog. Different kind of lines represent the consistency within 1-$\sigma$ (solid lines and filled symbols) and 2-$\sigma$ (dotted lines and empty symbols). Results are grouped by planet multiplicities where the total amount of planets are labeled at bottom}.}
    \label{percentagess_fig}
\end{figure}

\rev{In the work carried out by \cite{2010ApJ...709L..44N} the authors test their tool \texttt{TTVIM} (which combines a minimization algorithm with an analytic approximation for inverting TTVs) using synthetic transit observations with the aim of recovering planet masses and orbital elements of a non-transiting planet. 
\cite{2010ApJ...709L..44N} used upper limits in relative and absolute errors in planet parameters to decide whether the solutions were correctly recovered or not. 
Unlike \cite{2010ApJ...709L..44N},} we considered as recovered to those solutions from the posterior distributions consistent with the $true$ values within 68\% (1-$\sigma$) and 95\% (2-$\sigma$) \revi{credible} interval of the total posteriors, assuming that they follow a normal distribution. 
\rev{These solutions are marked in Figure \ref{recovered} as circles and squares.}
The comparison is made for planet parameters: mass, period ($P$), eccentricity \revi{($ecc$)},  periastron longitude ($\varpi$; defined as the sum of ascending node and argument of periastron, $\varpi = \Omega + \omega$) and the true longitude ($\ell$;  defined as $ \ell = \nu + \varpi$, where $\nu$ is the true anomaly, which is obtained from a Fourier expansion with the terms of the mean anomaly and eccentricity).
\revi{Some of our posteriors exhibit more than one peak, manifesting the nature of the selected sampler. Median values and errors are usually centered on the prominent peaks so the data in Figure \ref{recovered} are representative of our results. In practice, a dynamical stability test could help to distinguish the physically possible solutions. This study is underway but it is out of the scope of the present work.}

\revi{From Figure \ref{recovered}, \rev{we take the discrete values of the differences between recovered and $true$ data} (excluding the non-recovered data marked as gray crosses)}, assuming that discrete medians are representative of the MCMC results. This was done for each planet multiplicity and planet parameter. 
From the resulting distributions, we measure the standard deviation that represent, on average, the typical precision achieved in the recovery test for different parameters and planet multiplicities.
For masses, we found the minimum dispersion of \revi{0.7}  $M_{\oplus}$ for the two-planet systems, and a maximum of \revi{14} $M_{\oplus}$ for five-planet systems. 
For $P$, a minimum dispersion of \revi{9} seconds for two-planets and a maximum of \revi{110} seconds for five-planet systems. 
For $ecc$, a minimum dispersion of  \revi{0.007} for two-planets, and a maximum of \revi{0.03 for three-planet} systems. 
For $\varpi$, a minimum of \revi{35$^{\circ}$ for five-planets}, and a maximum of \revi{50}$^{\circ}$ for \revi{three-planet} systems. 
For $\ell$, a minimum dispersion of \revi{3}$^{\circ}$ for two-planets, and a maximum of \revi{7.5}$^{\circ}$ for \revi{four-planet} systems.
Intermediate dispersions were found for multiplicities not mentioned in these ranges.

We also calculate the  Kolmogorov-Smirnov (KS) statistic over the whole distribution of $true$ values and the posterior medians for all the parameters, for planets in the full catalog. We found, for masses, a $KS$-statistic=\revi{0.07} ($p$-value=\revi{0.96}); for $P$, $KS$-statistic=0.01 ($p$-value=1.0); for $ecc$, $KS$-statistic=\revi{0.25} ($p$-value=\revi{3$\times10^{-3}$}); for $\varpi$ $KS$-statistic=\revi{0.17} ($p$-value=\revi{0.11}), and for $\ell$, $KS$-statistic=\revi{0.03} ($p$-value=0.99). Thus, we cannot reject the hypothesis that the distribution of masses, periods ($P$) and true longitudes ($\ell$) are statistically drawn from the same input distributions.

\revi{
Figure \ref{percentagess_fig} shows the percentage of planets consistent with the $true$ parameters within 1-$\sigma$ and 2-$\sigma$, when grouping planets with the same multiplicity. In Table \ref{table_percentages} we summarize these results also considering the global percentages for the full catalog (independently of multiplicity). 
For masses, periods and true longitudes, the global recovery percentages are consistent with the expected statistical values, \ie around $\sim$68\% of the planet parameters being recovered within 1-$\sigma$ and $\sim$95\% within 2-$\sigma$. These parameters also exhibit consistency with the input catalog according to the KS test. 
However, we note that $ecc$ and $\varpi$ have similar recovery percentages but far below these statistical expected values. For these two parameters we found respectively, 52\% and 55\% within 1-$\sigma$ and 80\% and 76\% within 2-$\sigma$ (see Table \ref{table_percentages}).

We investigate the possible causes of both, the similarity and low recovery rates by inspecting the dependence between both parameters. We found that $\sim$45\% of planets with $ecc\lesssim$0.05 recover both, $ecc$ and $\varpi$ at the same time. The remaining planets do not recover at least one of these parameters. In contrast, planets with $ecc\gtrsim$0.05 recover both parameters $\sim$90\% of the times, showing that the low recovery rates occur mainly for low eccentric orbits.
In our catalog, most of the planets belonging to systems with three or more planets have eccentricities below $\sim$0.05, as shown in Figure \ref{catalog_parameters}. Even more, from Figure \ref{percentagess_fig}, it is seen that eccentricity is the only parameter that diminishes its recovery rate as the number of planets increases (teal lines with stars). }

\revi{We find that the cause of the similarity and the global low recovery rate for $ecc$ and $\varpi$ is a combination of the difficulty to find a well-defined argument of periastron as the orbits tend to be more circular, which in our case occurs mainly for planets belonging to high multiplicity systems.
We will show in section \ref{dependence_recovery} that the trend of the eccentricity shown in Figure \ref{percentagess_fig} and the low recovery rate of $ecc$ (and consequently $\varpi$) can be partially explained by the diminishing of the number of transits available for external planets belonging to high multiplicity systems.
}

\subsection{TTVs}\label{sect_TTVs}

\begin{figure*}
\centering
\begin{tabular}{c}
  \includegraphics[scale=.47]{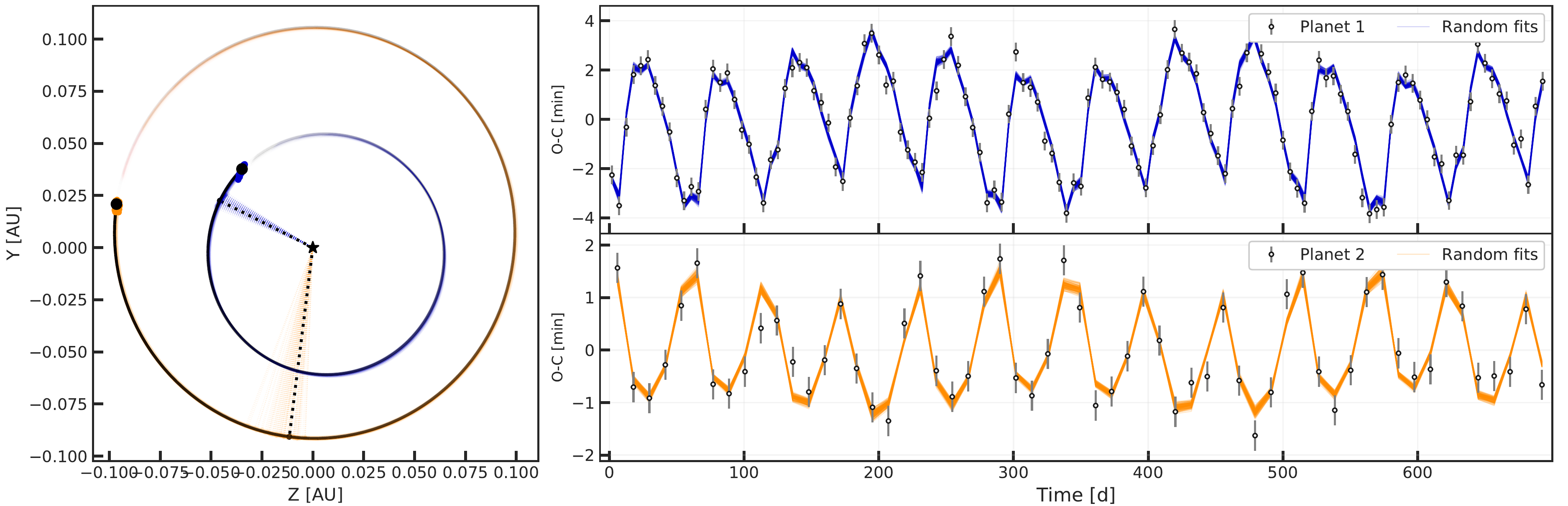} \\
  \includegraphics[scale=.47]{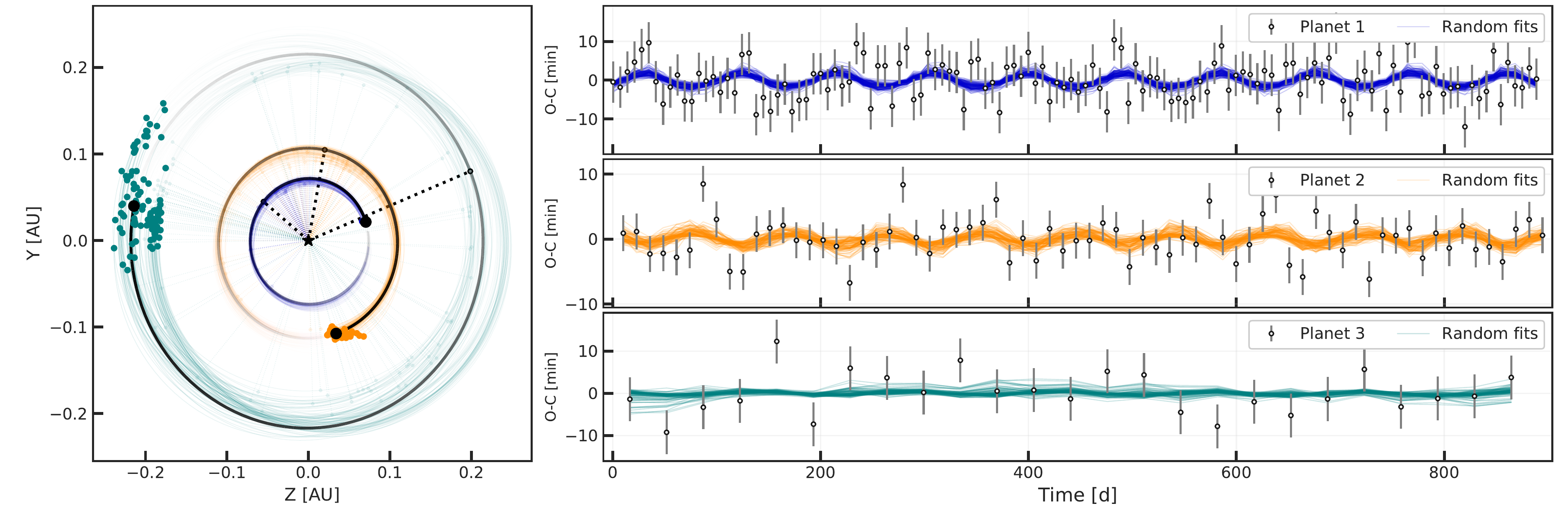} \\
\end{tabular}
\caption{Examples of TTVs fitting for a two-planet system (\revii{ID=pl2\_id52}; Table \ref{catalog2p}) and three planet system (\revii{ID=pl3\_id1}; Table \ref{catalog3p}). In left panels, dark solid lines correspond to the $true$ orbits and the lighter color orbits correspond to 100 orbital solutions randomly selected from the MCMC posteriors. Black dotted lines indicate \revi{the position of} the argument of periastron. In this reference frame, observer is located at +Z. \rev{These orbit plots have been made with an adapted version of the OrbitPlot in REBOUND \citep{2012A&A...537A.128R}.}  
\revi{Right panels show the simulated TTVs (circles with error bars), while colored lines show the 100 random fits with the same planet parameters of the orbits.} In both systems, the internal planets are labeled as Planet 1 and the order increases outwards.
}
\label{orbits_ttvs}
\end{figure*}

\begin{figure}[ht]
    \centering
    \includegraphics[scale=0.35]{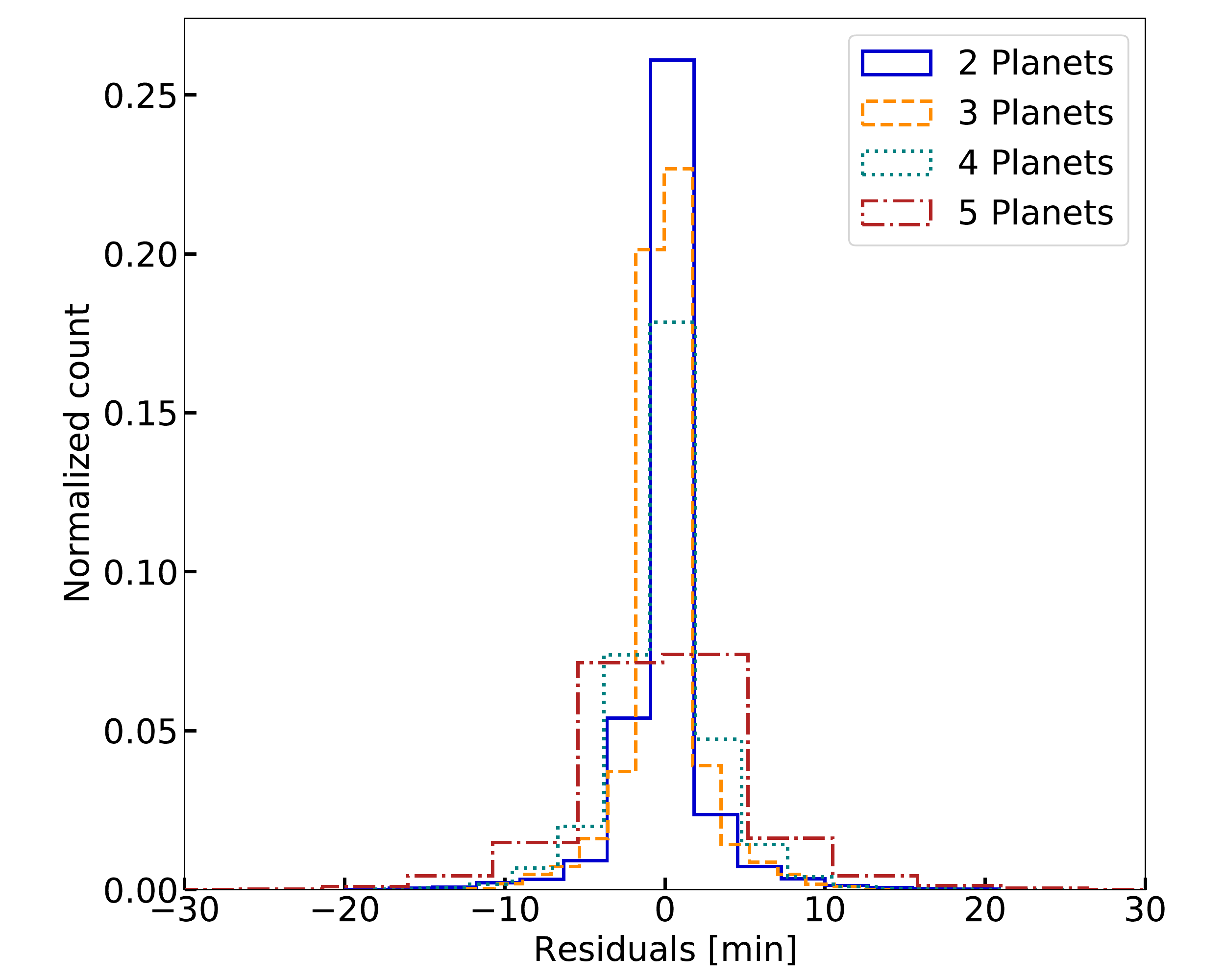}
    \caption{Typical residuals from the TTVs fitting grouped by planet multiplicity. Individual histograms were built by summing residuals of 100 random solutions per planet from the whole catalog.}
    \label{residuals}
\end{figure}

In Figure \ref{orbits_ttvs} we illustrate the TTVs of  two  systems with simulated data of different quality and their orbits. 
The top figure corresponds to the system with \revii{ID=pl2\_id52} in Table \ref{catalog2p} for which we estimate a signal-to-noise ratio (S/N) of \revi{8.4} and \revi{5.0} for planets 1 and 2, respectively. 
The bottom figure corresponds to the system with \revii{ID=pl3\_id1} in Table \ref{catalog3p} for which we estimate a S/N of \revi{1.6, 2.08 and 2.06} for planets 1, 2 and 3, respectively. 
\revi{Left panels show 100 orbital configurations (thin colored orbits) reconstructed from the planetary parameters taken randomly from our MCMC posteriors.} For comparison, the true orbits are plotted as solid black lines. \rev{Transits are detected each time the planets cross the +Z axis, which corresponds to the line of sight.}
\revi{Right panels show the 100 TTV signals (colored solid lines) using the same planet parameters of the orbits. Synthetic observations are shown by} empty circles with error bars. 
For these two systems, \revi{the TTV models are consistent with the synthetic data} but with a better planetary determination for the top system, which has a better S/N.
Hence, the quality in the transit times directly affects the determination of the \revi{orbital configurations}. For example, we  identify that in the case of the three-planet system, the argument of periastron show  \revi{a wide range of possible solutions}  with  respect  to  the  true  position (dashed  black  lines).

We statistically measured the fitted residuals from the whole catalog by taking 100 random samples per planet from the MCMC \revi{posteriors and then calculating} their corresponding TTVs. \revi{We take the differences between data and the fitted transit times, grouping these residuals according to their multiplicity. The resulting distributions are shown in Figure \ref{residuals}. 
We find typical standard deviation for these residuals being of \revi{2.5} minutes for two-planet, \revi{2.3} minutes for three-planet, \revi{3} minutes for four-planet and \revi{5.3} minutes for five-planet systems.}

\subsection{\revi{The impact of the data quality and quantity}}
\label{dependence_recovery}


\begin{figure*}[ht!]
\hspace*{-1cm}
\begin{tabular}{c}
  \includegraphics[scale=.37]{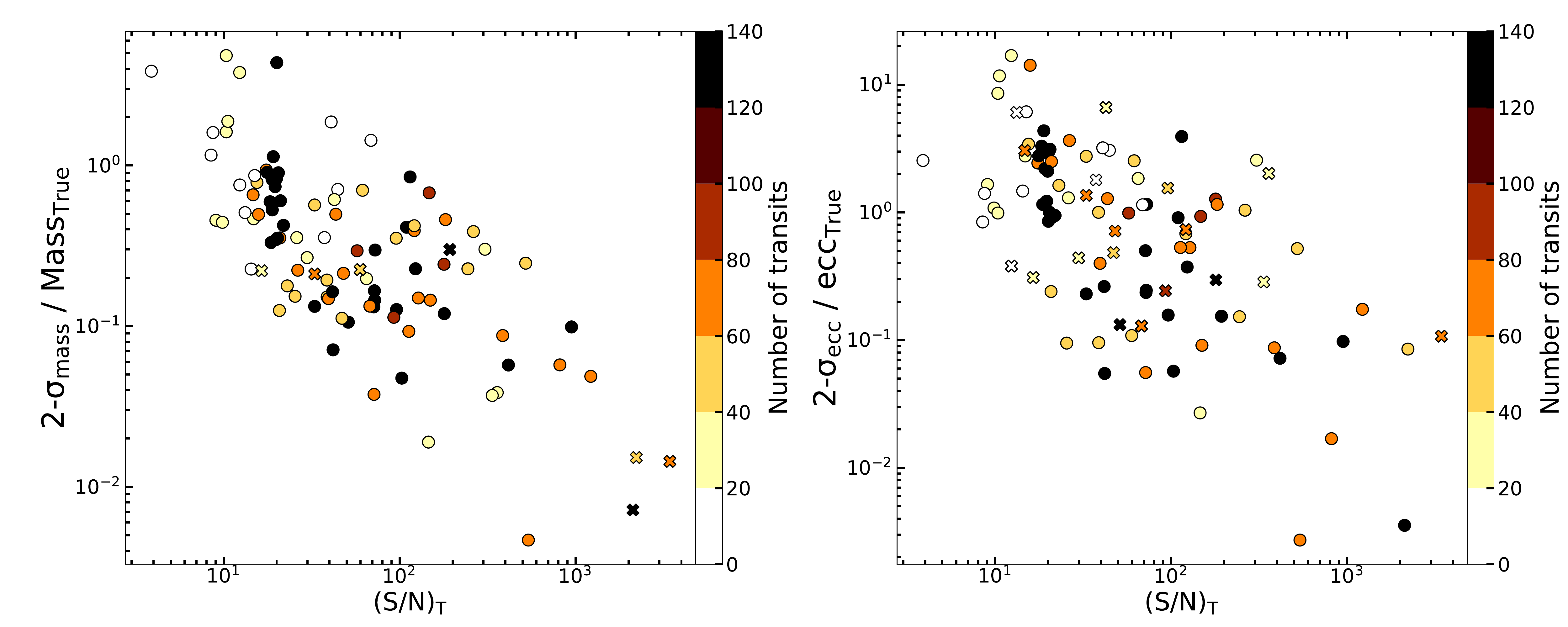} \\
  \includegraphics[scale=.37]{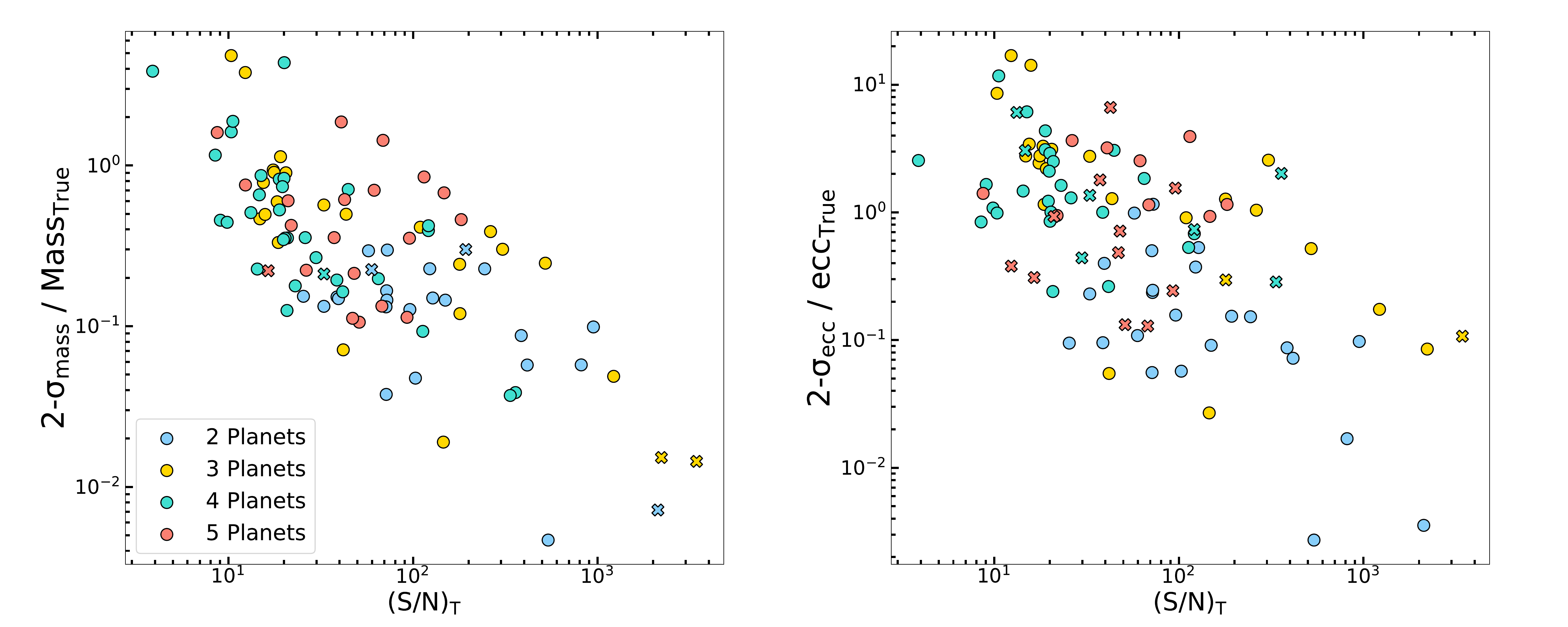}
\end{tabular}
\caption{Influence of the data quality and quantity over the proper determination of the planet mass (left panels) and eccentricity (right panels). Horizontal axes denote the \revii{total signal-to-noise $(S/N)_{T} = S/N\ \times \sqrt{\mathrm{N_{tran}}}$ } and vertical axes are the fractional errors derived from our results.
The 2-$\sigma$ corresponds to the 95\% \revi{credible} interval from the posteriors and $\mathrm{Mass_{True}}$ and $\mathrm{ecc_{True}}$ are the $true$ values in the catalog. The recovered planets are marked with circles and the non-recovered planets are marked with crosses. 
In top panels, colors correspond to the number of transits per planet ($\mathrm{N_{tran}}$) binned by \revi{20} transits. In the bottom panels, planets are colored according to the number of planets in their systems.
}
\label{sn_sigmas}
\end{figure*}

The TTVs inversion problem challenges not only the methods or algorithms, but also the observational data requirements.
In this subsection we address this issue by considering the quality (measured by the signal-to-noise ratio, S/N) and quantity ($\mathrm{N_{tran}}$) of the data required in order to correctly determine planetary parameters. We focus on the determination of planetary mass and eccentricity.

Previous works have proposed different answers regarding the required number of transits and data quality to invert TTVs. 
\cite{2019MNRAS.482.4965S} suggest that a large number of TTVs combined with a high S/N can be enough to robustly estimate planetary parameters without the necessity of radial velocity measurements.
A detailed analysis regarding the characterization of non-transiting planets was conducted by \citet{2011ApJ...727...74V}, finding that at least 50 transits are needed to invert the problem. 
\cite{2008ApJ...688..636N} developed a method based on perturbation theory to characterize two-planet systems, and found that a S/N $\sim$15-30 and about $\mathrm{N_{tran}}$ $\gtrsim$ 20  are typically required to uniquely characterize these systems.

We carried out an analysis of the parameters involved in the fitting procedure, looking for possible correlations between the data quality and quantity, the intrinsic planetary parameters (including the number of planets, $\mathrm{N_{pla}}$) and the results we found using \texttt{Nauyaca}.
\revii{A first inspection suggested a trend with the number of transits and thus it is convenient to define the total signal-to-noise $(S/N)_{T}$ as the product of $S/N\ \times \sqrt{\mathrm{N_{tran}}}$.}
Figure \ref{sn_sigmas} shows the fractional error in the determination of the planet mass and eccentricity,  and \revi{the dependency} with the \revii{$(S/N)_{T}$}, $\mathrm{N_{tran}}$ and  $\mathrm{N_{pla}}$.
Here,  2-$\sigma_{\mathrm{mass}}$ and 2-$\sigma_{\mathrm{ecc}}$ correspond to our derived uncertainties taken from the posteriors encompassing the 95\% \revi{credible} interval. 
These uncertainties are divided by the $true$ mass and eccentricity, and thus, Figure \ref{sn_sigmas} can work as a guide to put some constraints in the determination of these parameters \revi{given the quality and quantity of the data.}

In general, it is seen that the fractional error (related with the adequate determination of the parameters) have a correlation with the quality of the data and in second term with the quantity. Moreover, the deficiency of one of these quantities is in some cases compensated by the other one.
We observed that planets with individual high \revii{$(S/N)_{T}$} can constrain the planetary mass and eccentricity even with a relative low $\mathrm{N_{tran}}$. Complementary, a \revi{large} $\mathrm{N_{tran}}$ can compensate a low \revii{$(S/N)_{T}$}.
Planets with a low \revii{$(S/N)_{T}$} exhibit a larger dispersion of their fractional errors for their masses and eccentricities, although, those with a large $\mathrm{N_{tran}}$ usually have a better determination (\ie a lower fractional error).
Comparing the top and bottom panels, it can be distinguished that the majority of the non-recovered planets are those with a number of transits lower than $\lesssim$\revi{60}-80 \revi{(white/yellow crosses)} corresponding to systems with three, four and five planets. 
However, planets of these multiplicities with low \revii{$(S/N)_{T}$} can still find the correct solution but with a poor constraint \revi{(white/yellow circles in the upper left part of the top diagrams)}. 
\revii{Also notice from the bottom panels that the trend  between the fractional errors and the total \revii{$(S/N)_{T}$} seems to hold when grouped by planet multiplicities.}

From these results it is seen that \revii{$(S/N)_{T}$} and $\mathrm{N_{tran}}$ have a significant impact in the determination of these parameters. The low recovery rate for $ecc$ discussed in section \ref{sec_mcmc} is then explained by the low \revii{$(S/N)_{T}$} of our synthetic data and the low number of transits per planet ($\lesssim$20-40) for systems with many planets. 
Low \revii{$(S/N)_{T}$} can be due to noisy measurements of the transit timing or due to low amplitudes of the TTVs signals. In the first case, the limitation is in part by the instruments and observational strategy. In the second case, low amplitudes correspond to planets in less perturbed (almost circular) orbits.  In this situation, inverting TTV signals result in less constrained orbits.

\section{Considerations and caveats}\label{caveats}

\rev{Here we highlight many of the considerations made in this work that could be relevant for the users of \texttt{Nauyaca}. We also point out many of the limits and caveats when using this tool. }

\begin{itemize}
    \item In this work we assume the continuous monitoring of the planet transits and thus, there is not data gaps in our ephemeris. In practice this could be unrealistic, specially for ground-based observations where telescope time is limited. Although, as discussed by \citet{2008ApJ...688..636N}, the number of transits is determinant in the planetary characterization rather than the time distribution.

    \item We remind that there is not a unique way of choosing the fine-tuning parameters for running the algorithms in \texttt{Nauyaca}, since it will depend on the problem to solve. Although, the parameters chosen here for synthetic systems can work as a guide for real planets.

    \item Regarding the usage of  \texttt{Nauyaca} with real systems, the user must be aware that stellar mass and radius are  well restricted to consider them as constants. In cases where one or both of these parameters exhibit large uncertainties, we suggest running many simulations with a grid of stellar parameters including the values with the lower and upper uncertainties. 

    \item \revi{We remind that \texttt{TTVFast} and therefore \texttt{Nauyaca} is adapted to deal with planets around single parent stars, and hence planets in circumbinary systems or other configurations are not allowed.}
    
    \item We also must take into account that the fitting is performed over the mid-transit times instead of the own TTVs. Since TTVs are the transit time signals in an O-C diagram, the calculated data depends on the number of epochs and the used method to make the linear regression over the transit times. Fitting to TTVs instead of transit times alone can result in an imprecise period determination of about 40 minutes for a $\sim$70 days period planet, as shown by \cite{2018A&A...620A..88C}. For that reason, we performed the fitting methods over the raw mid-transit times.

    \item The simulations are performed over a pre-fixed time span with a pre-selected timestep which by default is set to $P_1/30$, where $P_1$ is the period of the internal planet. 

    \item \rev{When computing the mid-transit times, the planet radius is not taken into account} to assess whether a transit occurs. Only the coordinates of the planet center are considered to determine if the planet transits. Thus, grazing transits  \citep[as for example in WASP-67b;][]{2014A&A...568A.127M} are currently not considered.

\end{itemize}

\bigskip

\section{Summary and conclusions} \label{sec:Summary}

In this work we present \texttt{Nauyaca}\footnote{\url{https://github.com/EliabCanul/nauyaca}}, a Python package that encompasses minimization routines (Optimization module) and a Markov \revi{chain} Monte Carlo method (MCMC module) exclusively adapted to find planet parameters (masses and orbital elements) that best reproduce the transit times based on numerical simulations. 
\revi{Even though the numerical method is more computationally expensive (compared to analytical approximations)}, is more suitable to address more general situations, such as considering many planets with varied orbital configurations.
\texttt{Nauyaca} requires transit ephemeris per planet and \rev{stellar mass and radius}. \revi{Additionally, any previous knowledge about validity ranges can be supplied in order to better constrain the  parameter space.}

\rev{Previous studies of transit-timing analysis have used synthetic data to test new techniques or to quantify the relation between properties of TTVs  and planetary parameters \citep[\eg][]{2008ApJ...688..636N,2010ApJ...718..543M,2011ApJ...727...74V}. 
However, in most cases these studies have been limited to study two-planet systems where one of the planets transits and the other acts as the perturber. 
Here, we analyze a large sample of synthetic transiting planets with planetary parameters based on the current planet data from Exoplanet Archive (see section \ref{catalog_creation}). 
For these planets we calculate the mid-transit times to use them as input to \texttt{Nauyaca} (section \ref{sec:Rec_results}).  This allows us to characterize the performance of \texttt{Nauyaca} by measuring the consistency rate betwen the catalog entries and the parameters determined by the tool.}

For all the systems, we run optimization algorithms to test the performance for many planet multiplicities (section \ref{sec_optimizers}). Optimizers take advantage of the low consumption time  in contrast with a full MCMC run and provides an overall outlook of the regions in the parameter space with higher probability (the planet boundaries were defined in section \ref{planet_boundaries}).
We find that the best performance is achieved for the two-planet systems for any dimension, and for higher multiplicities the results are varied.
\revii{The optimizers would define high density regions for the masses of the planets within a factor $\sim$\revi{2-6}, allowing to initialize the chains around 20\% of the real masses.}
Eccentricities are in general well restricted only for two-planet systems.
\revii{Nevertheless, the optimizers for higher multiplicities are not well suited to define high density regions where the MCMC chains would be initialized.
Orbital angles remain in general loosely constrained, except for two-planet systems.}

We drawn the initial walker population for the parallel-tempering MCMC, using the best <10\% of the solutions from the optimizer runs. 
We find a good agreement between the input parameters in the catalog and those found in the recovery test with the MCMC, at 2-$\sigma$ (section \ref{sec_mcmc}). 
\revi{The global recovery percentages for masses, periods and true longitudes are in concordance with the statistical expected values of $\sim$68\% of the planet parameters being recovered within 1-$\sigma$ and $\sim$95\% within 2-$\sigma$. These parameters also exhibit consistency with the input catalog according to a Kolmogorov-Smirnov test. 
In contrast, eccentricities and periastron longitudes have similar low recovery rates. Even more, the recovery rate for eccentricity diminish as the number of planets increases.
We found that the cause of the similarity and the global low recovery rate between $ecc$ and $\varpi$  is a combination of the difficulty to determine the argument of periastron as the orbits tend to be circular and the reduction of the number of transits for external planets, mainly for high multiplicity systems. In these scenarios, the signal-to-noise ratio of the data plays a determinant role to correctly determine these parameters.

Depending on the planet multiplicities, typical mass precision accomplished in the recovery test range from \revi{$\sim$1-14} $\mathrm{M_{\oplus}}$. Periods achieve a precision between \revi{10 seconds} and \revi{2 minutes}. Eccentricities reach a precision between $\sim$0.01-0.03. Periastron longitudes and true longitudes have typical precisions of $\sim40^{\circ}$ and $6^{\circ}$, respectively.}

\rev{We investigate the dependence of data quality and quantity (section \ref{dependence_recovery}) and the proper determination of the planetary mass and eccentricity. We find that, in general, quality is better than quantity, although in many cases one parameter can compensate the deficiency of the other one. However, we warn that the results in this part of our study should be taken as a guide, since the simulated planet sample in our catalog is limited.}

Finally, we make suggestions about the fine-tuning parameters involved in the procedure. Depending on the computation facilities, parameters in Table \ref{table_parameters} could be a reasonable starting point to make optimization and MCMC runs. \rev{Note that in our mock catalog, the ``observed'' timespan and the number of transits in the systems are varied. Therefore, the fine-tuning parameters can be scaled to be adapted to specific problems.}
We suggest performing, at least, between 100-150 optimizer runs (N$_{\mathrm{opt}}$) times the number of planets in the system, and taking < 10-30\% of the best solutions to initialize walkers. The election of the \revi{initialization strategy presented in this work (section \ref{initialization_selection})} depends on the parameter space itself, which is unknown by nature. We suggest using the $ladder$ or $picked$ strategy if nothing is known about the parameter space (as in the majority of the situations) and the $gaussian$ \revi{strategy} if the parameter space is somehow constrained, as for example when considering fixed angles, fixed periods, circular orbits, etc. 
\revi{Of course, the suggested methodology used in this work is not mandatory when using \texttt{Nauyaca}, since for example, optimization routines and the proposed initialization strategies are optional. The user has the freedom to select the tools that are best suited to its TTVs inversion problem.  
However, we note an improvement in the MCMC performance and the results when following the proposed method shown in this work.}

In a forthcoming work, beyond the scope of this paper, we will show the application of the tool to more specific situations, as for example systems with non-transiting planets, with highly mutual inclinations and with missed transit data. We will also show the application to real systems in order to revisit planet parameters of previous characterized planets and also with new planets.

\revi{We provide the data used throughout this work in electronic format at \href{https://doi.org/10.5281/zenodo.5218498}{10.5281/zenodo.5218498}. \texttt{Nauyaca} first release can be found at \href{https://doi.org/10.5281/zenodo.5230451}{10.5281/zenodo.5230451}.
}


\section*{acknowledgments}

The authors gratefully acknowledge the computing time granted by DGTIC-UNAM for access to the Supercomputer Miztli in the HTC group, under the project with code LANCAD-UNAM-DGTIC-361. 
We thank Gabriel Perren and Luis M. Pavón for useful discussions.
\revi{We also thank the anonymous referees for their useful comments and suggestions made to improve the quality of the manuscript.}
EFCC also acknowledge for the PhD grant awarded by CONACYT Graduate Fellowship.
This work was supported by UNAM-PAPIIT IN-107518 and BG-101321.
HV was supported by the project UNAM-PAPIIT IN-101918. 
This research has made use of the NASA Exoplanet Archive, which is operated by the California Institute of Technology, under contract with the National Aeronautics and Space Administration under the Exoplanet Exploration Program.

\vspace{5mm}
\facilities{Supercomputer Miztli, Exoplanet Archive.}


\software{Numpy \citep{oliphant2006guide}, 
          Scipy \citep{2020SciPy-NMeth}, 
          H5py \citep{collette_python_hdf5_2014},
          Matplotlib \citep{Hunter:2007}, 
          Seaborn \citep{michael_waskom_2017_883859}.
          }

\appendix

\section{Catalog Tables}\label{catalog_tables}

Here we collect the data tables of the mock catalog that show the planet properties of the synthetic planetary systems. 
Tables are grouped by planet multiplicity. 
\revii{System IDs follow the syntax pl\{$\mathrm{N_{pla}}$\}\_id\{ID\}, where $\mathrm{N_{pla}}$ is the number of planets in the system and ID is an internal identifier number.}
Orbital elements correspond to the osculating elements at the simulated reference time of $t_0 = 0$ days. These tables can be found in electronic format at: \revi{\href{https://doi.org/10.5281/zenodo.5218498}{10.5281/zenodo.5218498}}.


\begin{deluxetable*}{cccccccccccc}[hbt!]
\tablecaption{Input catalog of two-planet systems. Column names are: system ID, stellar mass, stellar radius, planet number, planetary mass, period, eccentricity, inclination, argument of periastron, mean anomaly, ascending node and number of simulated transits. 
\label{catalog2p}}
\tablehead{
\colhead{System ID}  & \colhead{M$_{*}$} & \colhead{R$_{*}$} & \colhead{Planet} & \colhead{mass} & \colhead{$P$} & \colhead{$ecc$} & \colhead{$inc$} & \colhead{$\omega$} & \colhead{$M$} & \colhead{$\Omega$} & \colhead{N$_{\mathrm{tran}}$} \\
& \colhead{[M$_{\odot}$]} & \colhead{[R$_{\odot}$]} & & \colhead{[M$_{\oplus}$]} & \colhead{[d]} & \colhead{} & \colhead{[deg]} & \colhead{[deg]} & \colhead{[deg]} & \colhead{[deg]} & \colhead{}
}
\startdata 
pl2\_id0 & 0.91 & 0.86 & 1 & 5.721 & 5.1118 & 0.0921 & 91.264 & 154.81 & 147.77 & 89.75 & 130 \\
         &    &   & 2 & 27.969 & 11.762 & 0.1019 & 89.759 & 57.27 & 313.49 & 89.94 & 57 \\
pl2\_id1 & 1.09 & 1.3 & 1 & 12.599 & 9.55216 & 0.12 & 89.816 & 188.41 & 193.31 & 89.72 & 130 \\
         &    &   & 2 & 15.199 & 21.05799 & 0.0939 & 90.213 & 41.55 & 201.25 & 90.83 & 59 \\
pl2\_id7 & 1.09 & 1.61 & 1 & 0.769 & 15.09222 & 0.1012 & 90.244 & 280.58 & 160.98 & 90.04 & 131 \\
         &    &   & 2 & 0.899 & 22.79987 & 0.0895 & 89.978 & 338.53 & 186.65 & 91.98 & 86 \\
pl2\_id9 & 1.28 & 1.98 & 1 & 37.186 & 22.95311 & 0.0747 & 88.69 & 210.99 & 286.99 & 89.03 & 130 \\
         &    &   & 2 & 79.457 & 42.8698 & 0.1313 & 90.332 & 223.66 & 303.88 & 88.9 & 69 \\
pl2\_id23 & 1.0704 & 1.2077 & 1 & 11.592 & 10.68521 & 0.1346 & 89.983 & 150.27 & 358.03 & 90.51 & 130 \\
         &    &   & 2 & 12.233 & 20.88751 & 0.1098 & 90.039 & 41.13 & 358.67 & 89.81 & 67 \\
pl2\_id28 & 0.9127 & 0.8627 & 1 & 5.674 & 5.31963 & 0.2359 & 88.165 & 110.47 & 136.63 & 88.59 & 130 \\
         &    &   & 2 & 27.034 & 11.81586 & 0.0713 & 90.249 & 184.33 & 289.6 & 88.64 & 58 \\
pl2\_id35 & 1.0843 & 1.3948 & 1 & 10.257 & 10.78381 & 0.0945 & 90.567 & 67.59 & 349.14 & 90.66 & 130 \\
         &    &   & 2 & 13.115 & 19.67467 & 0.21 & 89.874 & 44.17 & 128.06 & 89.76 & 71 \\
pl2\_id38 & 1.0408 & 1.0681 & 1 & 10.07 & 12.40501 & 0.1372 & 90.949 & 41.03 & 91.03 & 88.58 & 130 \\
         &    &   & 2 & 7.749 & 20.61714 & 0.0941 & 89.34 & 36.34 & 185.82 & 91.93 & 78 \\
pl2\_id39 & 0.6513 & 0.6236 & 1 & 5.744 & 4.01342 & 0.1685 & 91.171 & 101.62 & 57.74 & 91.32 & 130 \\
         &    &   & 2 & 18.837 & 8.29871 & 0.1206 & 89.692 & 68.22 & 15.26 & 88.9 & 63 \\
pl2\_id45 & 0.9331 & 0.9199 & 1 & 24.2 & 16.70586 & 0.0898 & 89.331 & 40.92 & 221.72 & 88.17 & 130 \\
         &    &   & 2 & 119.359 & 30.74872 & 0.0931 & 90.339 & 108.98 & 130.39 & 90.93 & 70 \\
pl2\_id52 & 0.9133 & 0.8633 & 1 & 5.665 & 5.35468 & 0.1237 & 90.717 & 296.22 & 16.39 & 88.36 & 130 \\
         &    &   & 2 & 26.856 & 11.83319 & 0.0727 & 90.052 & 187.26 & 86.62 & 88.95 & 59 \\
pl2\_id62 & 1.0801 & 1.3877 & 1 & 9.775 & 17.95549 & 0.0884 & 91.163 & 174.38 & 74.34 & 89.82 & 130 \\
         &    &   & 2 & 8.062 & 32.45119 & 0.0399 & 89.168 & 124.35 & 28.63 & 89.02 & 72 \\
\enddata
\tablenotetext{}{}
\end{deluxetable*}


\begin{deluxetable*}{cccccccccccc}[hbt!]
\tablecaption{Input catalog of three-planet systems. Column names are: system ID, stellar mass, stellar radius, planet number, planetary mass, period, eccentricity, inclination, argument of periastron, mean anomaly, ascending node and number of simulated transits. 
\label{catalog3p}}
\tablehead{
\colhead{System ID}  & \colhead{M$_{*}$} & \colhead{R$_{*}$} & \colhead{Planet} & \colhead{mass} & \colhead{$P$} & \colhead{$ecc$} & \colhead{$inc$} & \colhead{$\omega$} & \colhead{$M$} & \colhead{$\Omega$} & \colhead{N$_{\mathrm{tran}}$} \\
& \colhead{[M$_{\odot}$]} & \colhead{[R$_{\odot}$]} & & \colhead{[M$_{\oplus}$]} & \colhead{[d]} & \colhead{} & \colhead{[deg]} & \colhead{[deg]} & \colhead{[deg]} & \colhead{[deg]} & \colhead{}
}
\startdata
pl3\_id1 & 1.08 & 1.49 & 1 & 7.31 & 6.88701 & 0.0218 & 89.978 & 309.12 & 121.87 & 90.51 & 130 \\
   &    &    & 2 & 7.049 & 12.81716 & 0.0297 & 89.835 & 10.79 & 150.06 & 91.57 & 70 \\
   &    &    & 3 & 3.0 & 35.32724 & 0.0096 & 90.831 & 68.07 & 213.07 & 90.92 & 25 \\
pl3\_id3 & 0.97 & 1.11 & 1 & 6.992 & 3.50448 & 0.0269 & 89.909 & 267.55 & 132.89 & 89.7 & 130 \\
   &    &    & 2 & 17.163 & 7.6433 & 0.0278 & 90.043 & 318.49 & 168.74 & 89.57 & 59 \\
   &    &    & 3 & 16.527 & 14.85607 & 0.0084 & 89.971 & 70.05 & 279.07 & 89.76 & 31 \\
pl3\_id4 & 1.08 & 1.0 & 1 & 7.31 & 34.54958 & 0.0384 & 90.17 & 133.3 & 186.18 & 88.79 & 130 \\
   &    &    & 2 & 4.132 & 66.07813 & 0.0077 & 90.58 & 111.04 & 96.02 & 90.46 & 68 \\
   &    &    & 3 & 133.488 & 125.84333 & 0.0373 & 90.045 & 159.98 & 192.7 & 89.06 & 36 \\
pl3\_id7 & 1.04 & 0.94 & 1 & 2.225 & 45.15447 & 0.0155 & 89.55 & 87.22 & 251.23 & 90.94 & 130 \\
   &    &    & 2 & 4.132 & 85.30295 & 0.0322 & 90.459 & 49.2 & 12.57 & 91.31 & 69 \\
   &    &    & 3 & 7.628 & 130.21746 & 0.028 & 89.744 & 23.47 & 72.62 & 90.52 & 45 \\
pl3\_id9 & 0.695 & 0.7204 & 1 & 5.402 & 1.75109 & 0.0465 & 91.436 & 38.82 & 345.74 & 89.37 & 130 \\
   &    &    & 2 & 4.694 & 4.59028 & 0.0236 & 88.986 & 161.25 & 348.33 & 90.29 & 49 \\
   &    &    & 3 & 6.988 & 8.2469 & 0.0252 & 89.615 & 8.84 & 310.5 & 89.16 & 28 \\
pl3\_id10 & 0.5224 & 0.4423 & 1 & 0.077 & 10.4707 & 0.0406 & 89.809 & 292.62 & 246.14 & 88.3 & 130 \\
   &    &    & 2 & 1.98 & 14.10509 & 0.0244 & 89.909 & 239.76 & 191.31 & 88.46 & 97 \\
   &    &    & 3 & 0.67 & 23.57467 & 0.0306 & 90.336 & 82.64 & 149.24 & 88.51 & 58 \\
pl3\_id11 & 1.0658 & 1.4631 & 1 & 7.238 & 6.71494 & 0.0359 & 90.821 & 4.54 & 160.56 & 90.27 & 130 \\
   &    &    & 2 & 6.862 & 12.53776 & 0.0074 & 90.446 & 204.83 & 162.09 & 90.43 & 70 \\
   &    &    & 3 & 3.032 & 34.44641 & 0.0062 & 88.187 & 291.73 & 88.84 & 91.03 & 26 \\
pl3\_id16 & 0.7275 & 0.7804 & 1 & 4.713 & 2.82797 & 0.0348 & 90.742 & 65.05 & 64.67 & 90.01 & 130 \\
   &    &    & 2 & 1.662 & 5.08958 & 0.0397 & 89.0 & 17.85 & 26.57 & 88.51 & 73 \\
   &    &    & 3 & 4.073 & 7.75808 & 0.0164 & 89.538 & 350.87 & 3.59 & 88.92 & 48 \\
\enddata
\tablenotetext{}{}
\end{deluxetable*}


\begin{deluxetable*}{cccccccccccc}[hbt!]
\tablecaption{Input catalog of four-planet systems. Column names are: system ID, stellar mass, stellar radius, planet number, planetary mass, period, eccentricity, inclination, argument of periastron, mean anomaly, ascending node and number of simulated transits. 
\label{catalog4p}}
\tablehead{
\colhead{System ID}  & \colhead{M$_{*}$} & \colhead{R$_{*}$} & \colhead{Planet} & \colhead{mass} & \colhead{$P$} & \colhead{$ecc$} & \colhead{$inc$} & \colhead{$\omega$} & \colhead{$M$} & \colhead{$\Omega$} & \colhead{N$_{\mathrm{tran}}$} \\
& \colhead{[M$_{\odot}$]} & \colhead{[R$_{\odot}$]} & & \colhead{[M$_{\oplus}$]} & \colhead{[d]} & \colhead{} & \colhead{[deg]} & \colhead{[deg]} & \colhead{[deg]} & \colhead{[deg]} & \colhead{}
}
\startdata
pl4\_id0 & 0.69 & 0.7 & 1 & 11.267 & 0.65853 & 0.042 & 90.167 & 194.71 & 163.89 & 90.11 & 130 \\
   &    &    & 2 & 0.289 & 7.81455 & 0.051 & 90.613 & 159.71 & 101.81 & 88.59 & 11 \\
   &    &    & 3 & 8.899 & 14.70854 & 0.0327 & 90.045 & 181.81 & 76.05 & 89.8 & 6 \\
   &    &    & 4 & 14.299 & 19.46914 & 0.0082 & 89.852 & 117.87 & 82.94 & 90.06 & 4 \\
pl4\_id2 & 1.28 & 1.52 & 1 & 10.488 & 3.74307 & 0.0287 & 90.513 & 314.46 & 190.51 & 88.26 & 130 \\
   &    &    & 2 & 15.574 & 10.42773 & 0.0403 & 89.628 & 281.5 & 301.3 & 88.81 & 47 \\
   &    &    & 3 & 106.155 & 22.34794 & 0.0147 & 90.044 & 310.83 & 231.24 & 88.63 & 22 \\
   &    &    & 4 & 34.961 & 54.40659 & 0.0229 & 90.077 & 235.71 & 31.83 & 88.83 & 9 \\
pl4\_id3 & 1.0 & 1.04 & 1 & 5.301 & 6.16484 & 0.0177 & 90.556 & 47.02 & 108.04 & 90.16 & 130 \\
   &    &    & 2 & 10.488 & 13.56747 & 0.0189 & 89.597 & 194.95 & 171.7 & 90.32 & 59 \\
   &    &    & 3 & 8.101 & 23.97665 & 0.0165 & 90.391 & 207.63 & 258.73 & 90.89 & 33 \\
   &    &    & 4 & 11.124 & 43.86076 & 0.009 & 89.728 & 73.95 & 273.67 & 90.43 & 18 \\
pl4\_id7 & 1.028 & 1.0878 & 1 & 5.092 & 5.81643 & 0.0334 & 88.339 & 49.89 & 70.58 & 87.76 & 130 \\
   &    &    & 2 & 10.36 & 12.55872 & 0.0189 & 89.747 & 175.26 & 215.81 & 89.24 & 61 \\
   &    &    & 3 & 7.6 & 22.10718 & 0.0228 & 90.328 & 67.44 & 114.15 & 90.16 & 34 \\
   &    &    & 4 & 10.83 & 40.44915 & 0.0403 & 90.276 & 216.75 & 198.03 & 90.8 & 19 \\
pl4\_id8 & 1.0624 & 1.3935 & 1 & 6.392 & 6.79961 & 0.0573 & 89.9 & 337.25 & 301.98 & 89.45 & 130 \\
   &    &    & 2 & 7.688 & 11.63317 & 0.0116 & 88.195 & 158.65 & 100.6 & 90.52 & 76 \\
   &    &    & 3 & 8.049 & 19.20383 & 0.0272 & 91.121 & 26.99 & 166.45 & 91.86 & 46 \\
   &    &    & 4 & 7.836 & 31.30115 & 0.0293 & 89.897 & 125.59 & 297.78 & 90.03 & 29 \\
pl4\_id9 & 1.1894 & 1.3388 & 1 & 10.829 & 11.76765 & 0.0486 & 90.31 & 210.63 & 299.0 & 91.07 & 130 \\
   &    &    & 2 & 7.719 & 24.42051 & 0.0079 & 90.433 & 120.73 & 69.75 & 89.35 & 62 \\
   &    &    & 3 & 23.683 & 46.86163 & 0.0169 & 89.907 & 300.94 & 231.49 & 89.62 & 32 \\
   &    &    & 4 & 9.565 & 76.28288 & 0.0666 & 89.866 & 162.63 & 222.65 & 90.03 & 20 \\
pl4\_id11 & 1.0606 & 1.3832 & 1 & 6.36 & 6.78081 & 0.0352 & 90.006 & 198.33 & 243.94 & 88.32 & 130 \\
   &    &    & 2 & 7.77 & 11.68647 & 0.0237 & 89.284 & 163.69 & 31.72 & 87.64 & 75 \\
   &    &    & 3 & 8.05 & 19.35073 & 0.0635 & 89.77 & 249.47 & 31.64 & 89.32 & 46 \\
   &    &    & 4 & 7.932 & 31.66494 & 0.0158 & 90.597 & 250.34 & 179.08 & 90.15 & 28 \\
pl4\_id13 & 1.1416 & 1.2819 & 1 & 4.244 & 4.40357 & 0.0435 & 88.887 & 76.63 & 318.04 & 90.25 & 130 \\
   &    &    & 2 & 9.84 & 8.45655 & 0.0132 & 88.65 & 63.97 & 97.3 & 88.76 & 67 \\
   &    &    & 3 & 5.565 & 14.52725 & 0.0411 & 90.086 & 172.97 & 94.37 & 90.04 & 39 \\
   &    &    & 4 & 9.635 & 26.67856 & 0.0051 & 91.074 & 299.01 & 232.92 & 89.57 & 21 \\
\enddata
\tablenotetext{}{}
\end{deluxetable*}


\begin{deluxetable*}{cccccccccccc}[hbt!]
\tablecaption{Input catalog of five-planet systems. Column names are: system ID, stellar mass, stellar radius, planet number, planetary mass, period, eccentricity, inclination, argument of periastron, mean anomaly, ascending node and number of simulated transits. 
\label{catalog5p}}
\tablehead{
\colhead{System ID}  & \colhead{M$_{*}$} & \colhead{R$_{*}$} & \colhead{Planet} & \colhead{mass} & \colhead{$P$} & \colhead{$ecc$} & \colhead{$inc$} & \colhead{$\omega$} & \colhead{$M$} & \colhead{$\Omega$} & \colhead{N$_{\mathrm{tran}}$} \\
& \colhead{[M$_{\odot}$]} & \colhead{[R$_{\odot}$]} & & \colhead{[M$_{\oplus}$]} & \colhead{[d]} & \colhead{} & \colhead{[deg]} & \colhead{[deg]} & \colhead{[deg]} & \colhead{[deg]} & \colhead{}
}
\startdata
pl5\_id0 & 0.81 & 0.76 & 1 & 4.3 & 5.28442 & 0.0286 & 87.353 & 247.95 & 146.13 & 89.42 & 130 \\
   &    &    & 2 & 3.0 & 7.0787 & 0.0118 & 89.263 & 284.41 & 138.07 & 89.43 & 98 \\
   &    &    & 3 & 3.814 & 10.31011 & 0.0504 & 91.07 & 50.12 & 206.02 & 90.19 & 67 \\
   &    &    & 4 & 8.899 & 16.1445 & 0.0194 & 91.287 & 113.77 & 39.55 & 89.8 & 42 \\
   &    &    & 5 & 5.2 & 27.4549 & 0.0373 & 89.23 & 338.07 & 141.81 & 90.33 & 25 \\
pl5\_id1 & 0.69 & 0.64 & 1 & 9.535 & 5.71468 & 0.0502 & 90.721 & 290.05 & 321.9 & 89.96 & 130 \\
   &    &    & 2 & 4.132 & 12.44489 & 0.0275 & 89.508 & 253.55 & 49.09 & 89.22 & 60 \\
   &    &    & 3 & 13.984 & 18.15237 & 0.0293 & 89.873 & 283.18 & 204.01 & 89.39 & 41 \\
   &    &    & 4 & 35.915 & 122.39306 & 0.0341 & 89.783 & 18.1 & 213.39 & 90.14 & 6 \\
   &    &    & 5 & 34.961 & 267.23151 & 0.0313 & 89.853 & 155.82 & 85.67 & 90.09 & 3 \\
pl5\_id2 & 0.8053 & 0.7553 & 1 & 4.504 & 5.31476 & 0.0202 & 88.981 & 342.17 & 193.54 & 89.05 & 130 \\
   &    &    & 2 & 3.044 & 7.24149 & 0.0787 & 90.105 & 74.54 & 256.88 & 90.4 & 95 \\
   &    &    & 3 & 4.21 & 10.63179 & 0.0339 & 90.111 & 290.26 & 238.28 & 89.61 & 65 \\
   &    &    & 4 & 9.951 & 20.27765 & 0.0031 & 89.968 & 175.85 & 193.59 & 89.92 & 34 \\
   &    &    & 5 & 6.358 & 36.78033 & 0.0299 & 90.235 & 264.02 & 264.33 & 88.03 & 18 \\
pl5\_id3 & 0.6966 & 0.6466 & 1 & 9.245 & 5.69128 & 0.0371 & 90.184 & 288.33 & 278.12 & 89.56 & 130 \\
   &    &    & 2 & 4.069 & 12.14691 & 0.0362 & 90.147 & 301.5 & 328.87 & 89.41 & 61 \\
   &    &    & 3 & 13.421 & 17.72588 & 0.0189 & 90.025 & 355.59 & 331.8 & 89.31 & 42 \\
   &    &    & 4 & 34.419 & 116.48534 & 0.0301 & 89.696 & 161.07 & 110.69 & 90.03 & 6 \\
   &    &    & 5 & 33.314 & 253.77612 & 0.0382 & 90.213 & 221.67 & 152.91 & 90.3 & 3 \\
\enddata
\tablenotetext{}{}
\end{deluxetable*}


\bibliographystyle{yahapj}
\bibliography{Bibliography2}

\end{document}